\documentclass[12pt]{article}
\usepackage{amsmath,amssymb,epsfig}
\begin{document}
\parindent=0pt
\parskip=6pt
\rm

\begin{center}
{\bf \Large Fluctuation and gauge effects on the critical behavior of
 superconductors}

\vspace{0.1cm}
 Diana V. SHOPOVA and Dimo I.
 UZUNOV

{\em  CP Laboratory, Institute of Solid State Physics,
 Bulgarian Academy of Sciences, BG-1784 Sofia, Bulgaria, and \\
 Max-Plank-Institut  f\"{u}r Physik komplexer Systeme,
N\"{o}thnitzer Str. 38, 01187 Dresden, Germany.}
\end{center}

$^{\ast}$ Corresponding author: sho@issp.bas.bg

{\bf Key words}: thermodynamics, order parameter, specific heat,
equation of state, renormalization group, fluctuation.

\vspace{0.2cm}

{\bf PACS}: 74.20.De, 74.78.Db

\vspace{0.2cm}

\begin{abstract}

Gauge effects on the fluctuation properties of the
normal-to-superconducting phase transition in bulk and thin film
superconductors are reviewed. Similar problems in the description
of other natural systems (liquid crystals, quantum field theory,
early universe) are also discussed. The relatively strong gauge
effects on the fluctuations of the ordering field at low spatial
dimensionality $D$ and, in particular, in thin (quasi-2D) films
are considered in details. A special attention is paid to the
fluctuations of the gauge field. It is shown that the mechanism,
in which these gauge fluctuations affect the phase transition
order and other phase transition properties varies with the
variation of spatial dimensionality $D$. The problem for the
experimental confirmation of theoretical predictions about the
order of phase transitions in gauge systems is discussed. Related
topics: gauge effects on the critical behavior of unconventional
superconductors, disorder, quantum fluctuations in a close
vicinity of ultra-low phase transition temperatures , are also
briefly discussed.
\end{abstract}

\section{INTRODUCTION}

\subsection{ Gauge effects and superconductivity}

A remarkable example of gauge theory in condensed matter physics
is the Ginzburg-Landau (GL) functional of
superconductivity~\cite{Ginzburg:1950, Lifshitz:1980}. The latter
is invariant towards both global $U(1)$ rotations of the order
parameter field $\psi(\vec{x})$ and local gauge transformations
(rotations) of the same field and the vector potential
$\vec{A}(\vec{x})$ of magnetic induction $\vec{B}$. These two
properties of gauge invariance define the global and local $U$(1)
symmetries of the GL theory. In both cases the gauge group is a
one-dimensional Abelian continuous group, $U$(1). The spontaneous
breaking of these global and local symmetries below the phase
transition point allows the appearance of the superconducting
phases: the uniform Meissner phase and, under certain
circumstances, the mixed (Abrikosov vortex~\cite{Abrikosov:1957})
phase. While the Meissner phase [$\langle\psi\rangle \neq 0 $,
$\langle\vec{B}\rangle =0$] is a mere product of the global $U$(1)
symmetry breaking, the vortex phase [$\langle\psi(\vec{x})\rangle
\neq 0 $, $\langle \vec{B}(\vec{x}) \rangle \neq 0$] , where the
equilibrium field configurations of both order parameter field
$\psi(\vec{x})$ and magnetic induction $\vec{B}(\vec{x})$ are
spatially nonuniform, occurs as a result of the spontaneous
breaking of local $U$(1) symmetry. The local gauge is important
also for the description of magnetic field penetration in both
Meissner and vortex phases. The penetration is described by an
additional characteristic length, the London penetration
length~\cite{Ginzburg:1950,Lifshitz:1980}, which is different from
zero only when the symmetry is broken, i.e., in the ordered
phases, where the equilibrium value of $\psi$ is different from
zero.  The local $U$(1) symmetry is present when the
superconductor is in external magnetic field, or, when this field
is equal to zero but magnetic fluctuations exist and determine the
vector potential $\vec{A}(\vec{x})$ as a purely fluctuating field.

The fluctuations of relevant fields, $\psi(\vec{x})$ and
$\vec{A}(\vec{x})$ in usual superconductors are small and can be
neglected. Therefore, the GL functional can be investigated in the
lowest order mean field (MF) approximation (alias, ``tree
approximation''~\cite{Uzunov:1993, Zinn-Justin:1993}). This is the
usual way of treatment of the GL free energy, in particular, the
study of GL equations~\cite{Ginzburg:1950, Lifshitz:1980,
Abrikosov:1957}. Within the tree approximation, the phase
transition from normal to superconducting state in a zero magnetic
field is of second order~\cite{Ginzburg:1950, Lifshitz:1980,
Uzunov:1993}.

For long time this phase transition has been considered as one of
the best examples of second order phase transitions, which has an
excellent description within MF. But in 1974 Halperin, Lubensky
and Ma (HLM)~\cite{Halperin:1974} showed that the magnetic
fluctuations change the order of the superconducting phase
transition in a zero external magnetic field ($H_0 = |\vec{H}_0| =
0)$), i.e., the order of the phase transition from
normal-to-uniform (Meissner) superconducting state at $T_{c0} =
T_c(H_0 = 0)$ (see, also Ref.~\cite{Chen:1978}). Since then this
fluctuation change of the order of normal-to-superconducting phase
transition ({\em HLM effect}) has been under debate. After 1974 an
overwhelming amount of theoretical research on this topic has been
performed but up to now there is no complete consensus about the
order of this phase transition.

In this review we shall consider some aspects of this problem but
we shall not be able to discuss or mention all relevant
contributions. However, it is important to emphasize that the
investigation of the properties of the superconducting phase
transition is also important to other areas of physics. Here we
shall briefly enumerate and discuss several examples.

 \subsection{Quantum field theory and other problems}

 In elementary particle physics the gauge invariant theory
similar to the GL theory of superconductivity is called Abelian
Higgs Model~\cite{Higgs:1964,
Guralnik:1964,Englert:1964,Kibble:1967, Coleman:1985,
Guidry:2004}. The same global $U$(1) and local $U$(1) gauge
symmetries are present there but the phenomenon of spontaneous
symmetry breaking occurs only for imaginary mass of the Higgs
($\psi-$) field, which is the analog of the field $\psi$,
describing the Cooper pair in a superconductor. In the absence of
spontaneous symmetry breaking this model would describe an
ordinary electrodynamics of charged scalars, but the situation
becomes more interesting when the mentioned mass is imaginary and
the breaking of symmetry occurs. Now the symmetry breaking
phenomena receive other names and another physical interpretation.
The breaking of both global and local gauge symmetries ensures a
mechanism for a transformation of the two initial scalar fields --
analogs of the components $\psi^{\prime}$ and $\psi^{\prime
\prime}$ of the complex field $\psi = \psi^{\prime} +
i\psi^{\prime \prime}$, and two massless photon fields -- analogs
of the two independent components $A_j$  of the vector potential
$\vec{A}=\{A_j; j=1,2,3; \nabla . \vec{A} =0\}$ in a
superconductor, to four massive particle fields: the so-called
Higgs boson, which is analog of the spontaneous order
 $|\psi| >0 $ in a
superconductor, and three massive vector field components, i.e. a
massive three dimensional vector field. The mass of this new
vector field is proportional to the electric charge $|q|$ and
magnitude $|\psi|$ of the Higgs field and, as a matter of fact,
this is exactly the way, in which the London penetration length in
a superconductor depends on the electron charge $|e|$ and the
modulus $|\psi|$ of the superconducting order parameter $\psi$.
Thus the spontaneous breaking of the local gauge symmetry leads to
the formation of massive particles without spoiling the gauge
invariance of the theory. This is called ``the Higgs mechanism''.
It plays a fundamental role in the unified theory of
electromagnetic, weak and strong interactions (see, e.g.,
Ref.~\cite{Coleman:1985, Guidry:2004}).

It is easy to see that there is something quite common between the
phenomena of spontaneous breaking of the continuous symmetries in
superconductivity theory and in quantum field theory. The
superconducting phase $|\psi| > 0$ is the exact analog of the
Higgs boson in the Abelian-Higgs model, whereas the appearance of
a massive vector field has its analog in the finite London
penetration length, mentioned above. Of course, there is no
obstacles in interpreting the phenomena of spontaneous symmetry
breaking in quantum field theory as phase transitions by taking
the Higgs mode mass as tuning parameter. The phase transition will
occur at zero mass of the Higgs boson.

A similar phenomenon of spontaneous breaking of both global $U$(1)
and local gauge symmetries is possible also within the scalar
electrodynamics due to mass insertions from the radiation
corrections, as shown in Ref.~\cite{Coleman:1973}. The radiation
corrections, in analogy with the magnetic fluctuations in a
superconductor, generate an imaginary mass to the initially
massless scalar field in this theory and the latter becomes very
similar to the Abelian-Higgs model. Here the symmetry breaking
leads to the appearance of massive scalar and vector fields
describing neutral scalar meson and vector meson,
respectively~\cite{Coleman:1973}. The radiation corrections to the
Lagrangian of the massless scalar
electrodynamics~\cite{Coleman:1973} resemble very much, in
particular in their mathematical form at $D=4$, the magnetic
fluctuation corrections to the GL free energy of 4D
superconductors~\cite{Chen:1978}. The interrelationships between
the superconductivity theory and the gauge theories of elementary
particles has been comprehensively discussed in
Ref.~\cite{Linde:1979}. Note also the relation between the
superconductivity GL functional and the $CP^{N-1}$ confinement
model (see Ref.~\cite{Hikami:1979}) and extensions to non-Abelian
theories~\cite{Brezin:1979}.

The gauge theories, mentioned so far, and their extensions have a
wide application in the description of the Early
Universe~\cite{Linde:1979, Vilenkin:1994}. Another interesting
gauge theory is that of the nematic-to-smectic {\em A} phase
transition in liquid crystals. According to the
Kobayashi-McMillan-de Gennes theory~\cite{ Kobayashi:1970,
McMillan:1972, Gennes:1972} the smectic-{\em A} order is described
by two order parameters: the nematic director vector and the
complex scalar which gives the long molecule centers of masses.
When a description, analogous to that of superconductors, is
introduced, as suggested by de Gennes~\cite{Gennes:1972}, the
director vector is substituted with a gauge vector field, which is
quite similar to the vector potential $\vec{A}$ in the GL
functional. Apart from some specific features intended to take
into account the liquid crystal anisotropy, the effective free
energy of the smectic A looks very much like the GL free energy of
superconductors (see, also, Refs.~\cite{Lubensky:1974,
Lubensky:1978, Gennes:1993}). A gauge theory in condensed matter
physics that has some (but not very close) similarity with the GL
free energy of superconductors is the Chern-Simons-Ginzburg-Landau
(CSGL) effective functional of the quantum Hall liquid state, in
particular, in the description of phase transitions between the
plateaus in the quantum Hall effect~\cite{Zhang:1989,
Schakel:1995, Wen:1993, Pryadko:1994}. Let us mention also the
liquid metallic hydrogen, where the problem of the
superconducting-to-superfluid phase transition~\cite{Babaev:2004}
is also related to the topics, discussed in this review.

\subsection{About the investigation of the fluctuation-driven
first order phase transition in superconductors}

Now we will focus on how the fluctuations effect the properties of
the normal-to-superconducting phase transition in a zero external
magnetic field, which is the subject of the present review and in
particular cases we shall refer to related topics in other natural
systems. We shall consider in more details the HLM effect in 2D
and quasi-2D superconductors with a special emphasis on the
problem  of the theoretical predictions reliability. Following
Ref.~\cite{Halperin:1974} we will use two theoretical methods.

In Sec.~II we use the MF like approximation of
Ref.~\cite{Halperin:1974},  where the spatial fluctuations of
superconducting order parameter $\psi$ are neglected, and the
magnetic fluctuations are taken into consideration. This treatment
was justified for well established type I superconductors, where
the London penetration length $\lambda$ is much smaller than the
coherence (correlation) length $\xi$. Due to the neglecting of
spatial fluctuations of $\psi$, the results of such MF treatment
should be valid outside the Ginzburg critical
region~\cite{Uzunov:1993}. A  weakly first order phase transition
occurs as a result of  new small $|\psi|^3$-term, which appears in
the effective free energy of 3D superconductors within the
framework of the MF like approximation~\cite{Halperin:1974}. But
this HLM effect was found to be very small and experimentally
unobservable even for well established type I bulk (3D-)
superconductors, such as Al, where the GL number $\kappa =
\lambda/\xi = 10^{-2}\ll 1$. It has been recently
shown~\cite{Folk:2001, Shopova:2002, Shopova2:2003, Shopova3:2003,
Shopova4:2003, Shopova5:2003, Todorov:2003} that the HLM effect is
much stronger in quasi-2D superconductors than in bulk (3D)
samples. Moreover, as shown in this series of papers, the effect
appears by a term of type $|\psi|^2\mbox{ln}|\psi|$ in the
effective free energy, and essentially depends on the thickness
$L_0$ of quasi-2D superconducting films. These circumstances
provide a real opportunity for an experimental verification of the
effect -- a topic of discussion throughout the present review, and
in particular, in Sec.~II.

We must emphasize that MF results are not valid for thermodynamic
 states in the Ginzburg critical region $(\delta T)_G
=|T_G-T_{c0}|/T_{c0}$ around the equilibrium phase transition
point $T_{c0}$. In certain classes of high-temperature
superconductors the Ginzburg region exceeds $0.1$~K whereas in
usual low-temperature superconductors it is very narrow, $(\delta
T)_G \sim 10^{-12} - 10^{-16}$~K, and the respective critical
 states are experimentally unaccessible (see, e.g.,~\cite{Uzunov:1993}).
If the metastability states, which go along with the weakly first
order phase transition predicted by MF, extend over temperature
intervals larger than the size of the Ginzburg region, one can
conclude that the MF prediction of the HLM effect is reliable. In
Sec.~II we justify  the MF analysis reliability  for element
superconductors as Al, W, In. We will show that the condition for
enough wideness of metastability regions is quite strong and,
perhaps, irrelevant in experiments. We will also demonstrate that
the magnitude of the critical magnetic field is crucial for HLM
effect to be observed in thin superconducting films (see also the
discussion in Sec.~II.9).

In Sec. III we review some results for the HLM effect obtained
with the help of the renormalization group (RG)
method~\cite{Uzunov:1993, Zinn-Justin:1993}. The latter allows
simultaneous treatment of both superconducting and magnetic
fluctuations in the asymptotically close vicinity of phase
transition point. The lack of fixed point of one-loop RG equations
for conventional superconductors in a zero external magnetic field
was interpreted as a signal for a fluctuation-driven first order
phase transition~\cite{Halperin:1974}. This result shows that the
local gauge magnetic fluctuations are relevant also in the
Ginzburg region of strong $\psi$-fluctuations, and under certain
circumstances, they can change the phase transition order to a
weakly first order, as is outside the critical region. The RG
investigations  of the superconducting phase transition order in a
zero magnetic field has been recently reviewed in
Ref.~\cite{Holovatch:1999}. While the latter review
 emphasizes the investigation of the HLM effect in high orders of
the loop expansion~\cite{Uzunov:1993, Zinn-Justin:1993}, here we
shall stress on RG results concerning the effects of anisotropy,
quenched disorder and extended symmetries of the  Higgs
$\psi$-field on the phase transition properties. The RG study of
unconventional superconductors will also be discussed
(Sec.~III.7). In the remainder of this paper we will consider
mainly the phase transition to the superconducting state but
results for other systems and related topics will be also
mentioned in brief; see, e.g., Sec.~III.3 -- III.6.

\section{MF STUDIES}

\subsection{General GL functional}

The GL free energy~\cite{Lifshitz:1980} of D-dimensional
superconductor of volume $V_D = (L_1...L_D)$ is given in the form
\begin{equation}
 F(\psi,\vec{A}) = \int d^D x \left[ a|\psi|^2 + \frac{b}{2}|\psi|^4 +
\frac{\hbar^2}{4m}\left|\left(\nabla - \frac{2ie}{\hbar
c}\vec{A}\right)\psi \right|^2 + \frac{\vec{B}^2}{8\pi}\right]\:.
\end{equation}
In Eq.~(1) the first Landau parameter $a = \alpha_0(T-T_{c0})$ is
expressed by the critical temperature $T_{c0}=T_c(H=0)$ in a zero
external magnetic field ($H=|\vec{H}|$), $b
> 0$ is the second Landau parameter and
$e \equiv |e|$ is the electron charge. The square $\vec{B}^2$ of
the magnetic induction $\vec{B} = (\vec{H} + 4\pi\vec{M})$, is
given by the vector potential $\vec{A}(\vec{x})=\{ A_j(\vec{x}),
\; j=1,...,D\}$ in the following way
\begin{equation}
\label{eq2}
 \vec{B}^2 = \frac{1}{2} \sum^{D}_{i,\:j\:=\:1} \left( \frac{\partial A_j}{\partial x_i} -
 \frac{\partial A_i}{\partial x_j}\right)^2\:,
\end{equation}
here the vector potential $\vec{A}(\vec{x})$ obeys the Coulomb
gauge $\nabla \cdot \vec{A}(\vec{x}) = 0$. For 3D superconductor
the relation $\vec{B} = \nabla \times \vec{A}(\vec{x})$ can be
used and when $\vec{B}=\vec{B}_0$ is uniform along the $z$- axis,
the Landau gauge $\vec{A}_0(\vec{x}) = B_0 (-y/2,-x/2,0)$ can be
applied. This representation can be generalized for $D >2$ -
dimensional systems, where the magnetic induction $B_0$ is a
second rank tensor~\cite{Lawrie1:1983}:
\begin{equation}
\label{eq3}
B_{0ij} =B_0 (\delta_{i1}\delta_{j2} -
\delta_{j2}\delta_{i1}).
\end{equation}

If we use the notation $\vec{x} = (x_1,x_2,\vec{r})$, where
$\vec{r}$ is a $(D-2)$ - dimensional vector, perpendicular to the
plane $(x_1,x_2)$, in the $3D$ case we have $\vec{r}=(0,0,z)$, and

\begin{equation}
\label{eq4}
B_j = \frac{1}{2}\epsilon_{jkl}B_{0kl} = B_0\delta_{j3}\;,
\end{equation}
where $\epsilon_{jkl}$ is the antisymmetric Levi-Civita symbol.
The Landau gauge and Eqs.~(\ref{eq3})~-~(\ref{eq4}) can be used
for uniform $\vec{B}=\vec{B}_0$ when $\delta \vec{B}$ -
fluctuations are neglected. In the prevailing part of our study we
will apply the general Coulomb gauge of the field
$\vec{A}(\vec{x})$, which does not exclude spatial dependent
magnetic fluctuations $\delta \vec{B}(\vec{x})$.

In nonmagnetic superconductors, where the mean value
$\langle\vec{M}\rangle =(\vec{M}-\delta\vec{M})$ of magnetization
$\vec{M}$ is equal to zero in the normal state in a zero external
magnetic field, the magnetic induction in presence of external
magnetic field takes the form:
\begin{equation}
\label{eq5} \vec{B} = \vec{H}_0 + \delta \vec{H}(\vec{x}) +4\pi\delta
\vec{M}(\vec{x})\;,
\end{equation}
where $\vec{H}_0$ is the (uniform) regular part of the external
magnetic field and $ \delta \vec{H}$ is an irregular part of
$\vec{H}$ created by uncontrollable  effects. We neglect the
irregular part $\delta \vec{H}$ and set $\vec{H}_0=0$, then
$\vec{B}$ contains only a fluctuation part $ \vec{B} \equiv \delta
\vec{B}(\vec{x}) = 4\pi\delta \vec{M}(\vec{x})$ that describes the
diamagnetic variations of $\vec{M}(\vec{x})$ around the zero value
$\langle\vec{M}\rangle =0$ due to fluctuations
$\delta\psi(\vec{x})$ of the ordering field $\psi(\vec{x})$ above
$(T>T_{c0})$ and below $(T<T_{c0})$ the normal-to-superconducting
transition at $T_{c0}$. Note, that the non-fluctuation  part
$\vec{A}_0=[\vec{A}(\vec{x})- \delta\vec{A}(\vec{x})]$ corresponds
to the regular part $\vec{B}_0 = (\vec{H}_0 + \langle
\vec{M}\rangle) = 0$ of $\vec{B}$ in nonmagnetic superconductors
$(\langle \vec{M}\rangle = 0)$ in a zero external magnetic field
$(\vec{H}_0 =0)$. Then we can set $\vec{A}_0(\vec{x})=0$ and,
hence, $\delta\vec{A}(\vec{x})=\vec{A}(\vec{x})$, so we have an
entirely fluctuation vector potential $\vec{A}(\vec{x})$, which
interacts with the order parameter $\psi(\vec{x})$. This
interaction can be of type $|\psi|^2A$ and  $|\psi|^2A^2$ and
generates all effects discussed in the paper.

We accept periodic boundary conditions for the superconductor
surface. This means to ignore the surface energy including the
additional energy due to the magnetic field penetration  in the
surface layer of thickness equal to the London penetration depth
$\lambda(T)=\lambda_0|t_0|^{-1/2}, \; t_0= |T-T_{c0}|/T_{c0}$;
$\lambda_0= (mc^2b/8\pi e^2\alpha_0 T_{c0})^{1/2}$ is the
``zero-temperature'' value of $\lambda$. This approximation is
adequate for superconductors of thickness $L_0 \gg \lambda(T)\gg
a_0$, where  $a_0$ is the lattice constant and $L_0 = min\{L_i,\:
i=1,...,D \}$. As we suppose the external magnetic field to be
zero $(H_0=0) $ or very small in real experiments, the requirement
$L_0 \gg \lambda(T)$ can be ignored and we have the simple
condition
 $L_0 \gg a_0$.

In microscopic models of periodic structures the periodic boundary
conditions confine the wave vectors $\vec{k}_i= \{k_i=(2\pi
n_i/L_i);\; i=1,...,D\}$ in the first Brillouin zone $[
-(\pi/a_0)\le k_i < (\pi/a_0)]$ and the expansion of their values
beyond this zone can be made either by neglecting the periodicity
of the crystal structure or on the basis of assumption  that large
wave numbers $k=|\vec{k}|$ have a negligible contribution to the
calculated quantities. The last argument is widely accepted in the
phase transition theory, where the long-wavelength limit $(ka_0\ll
1)$  can be used. In particular, this argument is valid in the
continuum limit $(V_D/a_0^D \to \infty)$. Therefore, for both
crystal and nonperiodic structures we can use the cutoff $\Lambda
\sim (\pi/a_0)$ and afterwards extend this cutoff to infinity,
provided the main contributions in the summations over $\vec{k}$
come from the relatively small wave numbers $(k \ll \Lambda)$. In
fact, this is a quasi-macroscopic description based on the GL
functional~(1), so the microscopic phenomena are excluded from our
consideration.

The GL free energy functional takes into account phenomena with
characteristic lengths $\xi_0$ and $\lambda_0$ or larger ($\xi$
and $\lambda$), where $\lambda(T)$ is the London penetration
length, mentioned above, and $\xi(T) = \xi_0|t|^{-1/2}$ is the
coherence length~\cite{Lifshitz:1980}; here $\xi_0 =(\hbar^2/4m
\alpha_0 T_{c0})^{1/2}$ is the zero-temperature coherence length.
In low-temperature superconductors $\xi_0$ and $\lambda_0$ are
much bigger than the lattice constant $a_0$.  Having in mind this
argument we will assume in our investigation  that $\Lambda \ll
(\pi/a_0)$. Whether the upper cutoff $\Lambda$ is chosen to be
either $\Lambda \sim 1/\xi_0$ or $\Lambda \sim 1/\lambda_0$ is a
problem that has to be solved by additional considerations.
According to arguments presented in Ref.~\cite{Folk:2001} and
Sec.~II.6, we will often make the choice $\lambda \sim
\xi_0^{-1}$.

We will use the Fourier expansion
\begin{equation}
\label{eq6} A_j(\vec{x}) = \frac{1}{V^{1/2}_D} \sum_k
A_j(\vec{k})e^{i \vec{k}. \vec{x}}
\end{equation}
and
\begin{equation}
\label{eq7} \psi (\vec{x}) = \frac{1}{V^{1/2}_D} \sum_k \psi
(\vec{k})e^{i \vec{k}. \vec{x}}\; ,
\end{equation}
where the Fourier amplitudes $A_j(\vec{k})$ obey the relation
$A_j^{\ast}(\vec{k}) = A_j(- \vec{k})$ and  $\vec{k} .
\vec{A}(\vec{k}) = 0$. The Fourier amplitude  $\psi (\vec{k})$ is
not equal to $\psi^{\ast} (- \vec{k})$ because $\psi (\vec{x})$ is
a complex function. For the same reason $\psi(0) \equiv \psi
(\vec{k} =0)$ is a complex number.

The functional (1) is invariant under global $U$(1) rotations,
defined by $\psi(\vec{x}) \rightarrow
\psi(\vec{x})\mbox{exp}(i\alpha)$, where the angle $\alpha$ does
not depend on the spatial vector $\vec{x}$, and under the local
$U$(1) gauge transformations $\psi(\vec{x}) \rightarrow
\psi(\vec{x})\mbox{exp}[i\alpha(\vec{x})]$, $\vec{A}(\vec{x})
\rightarrow \vec{A}(\vec{x}) + (\hbar c/2e)\vec{\nabla} \alpha
(\vec{x})$. According to the discussion in Sec.~I, we have to
investigate the spontaneous breaking of these symmetries, that is,
the ordered phases and the phase transitions in the
superconductor. The HLM effect,on which we are going to focus the
attention, is one of the results of the local $U$(1) symmetry
breaking.

\subsection{Notes about the MF like approximation}

The effect of the superconducting fluctuations $\delta \psi
(\vec{x})$ on the phase transition properties is very weak, as in
usual superconductors, and is restricted in a negligibly small
vicinity $(|t_0| \sim 10^{-12} \div 10^{-16})$ of  temperature
$T_{c0}$, we will assume that $\delta \psi (\vec{x}) = 0$, i.e.,
$\psi \approx \langle \psi (\vec{x}) \rangle$; from now on we will
denote $\langle \psi (\vec{x}) \rangle$ by $\psi$.
 Therefore, we apply the mean-field approximation
with respect to the order parameter $ \psi (\vec{x})$. Within this
approximation we will take into account the $\delta
\vec{A}(\vec{x})$-fluctuations of $\vec{B}_0 = 0$, i.e.,
$\vec{A}(\vec{x})= \delta \vec{A}(\vec{x})$. Furthermore, the
$\vec{A}(\vec{x})$-fluctuations can be integrated out from the
partition function, defined by:
\begin{equation}
\label{eq8} Z(\psi) = \int {\cal{D}}A e^{-F(\psi, \vec{A})/k_B T} \;,
\end{equation}
where the functional integral $ \int {\cal{D}}A$ is given by
\begin{equation}
\label{eq9} \int^{\infty}_ {-\infty} \prod_{j=1}^D \prod_{x \in V_D} d
A_j(\vec{x}) \delta[\mbox{div} \vec{A} (\vec{x})] \;.
\end{equation}
The integration is over all possible configurations of the field
$\vec{A}(\vec{x})$; the $\delta$-function takes into account the
Coulomb gauge.

The partition function $Z(\psi)$ corresponds to an effective free
energy $\cal{F}_D$ of the $D$-dimensional system:
\begin{equation}
\label{eq10} {\cal{F}}_D =  - k_B T \ln{Z(\psi)} .
\end{equation}
The magnetic fluctuations will be completely taken into account,
if only we are able to solve exactly the integral~(\ref{eq8}). The
exact solution can be done for a uniform order parameter $\psi$.
The uniform value of $\psi$ is different from the mean-field value
of $\psi$, because the uniform fluctuations of  $\psi(\vec{x})$
always exist, so we should choose one of these two
possibilities~\cite{Folk:2001, Shopova5:2003}. The problem of this
choice arises after calculating the integral~(\ref{eq8}) at the
next stage of consideration, when the effective free energy
$\cal{F}_D$ is analyzed and the properties of the superconducting
phase $(\psi > 0)$ are investigated. The effective free energy is
a particular case of the effective thermodynamic potential in the
phase transition theory~\cite{Uzunov:1993,Zinn-Justin:1993} and we
must treat the uniform $\psi$ in the way prescribed in the field
theory of phase transitions. It will become obvious from the next
discussion that we will use a loop-like expansion, which can be
exactly summed up to give a logarithmic dependence on $|\psi|$.

Due to the spontaneous symmetry breaking of the global $U$(1)
continuous symmetry of the ground state $\psi \neq 0$, the
effective free energies discussed in this Section depend on the
modulus $|\psi|$ of the complex number $\psi = |\psi| e^{i\theta}$
but not on the phase angle $\theta$, which remains arbitrary. That
is why we will consider the modulus $|\psi|$ as an ``effective
order parameter'' as the angle $\theta$ does not play any role in
the phenomena investigated in this Section. The quantity $|\psi|$
remains undetermined up to the stage when we determine the
equilibrium order parameter $|\psi_0|$ by the equation of state
$[\partial {\cal{F}}_D(\psi)/\partial \psi] = 0$. This equation
gives the equilibrium value $\psi_0$ of $\psi$ and the difference
$\delta \psi_0 = (\psi_0$ - $\psi)$ can be treated as the uniform
(zero dimensional) fluctuation of field $\psi(\vec{x})$. The
$\vec{x}$-dependent fluctuations $\delta \psi (\vec{x})$ have been
neglected because of the $\psi$ uniformity. The solution $\psi_0$
will be stable towards the uniform fluctuation $\delta \psi$,
provided the same solution $\psi_0 = |\psi_0|e^{i\theta_0}$
corresponds to a stable (normal or superconducting) phase; the
phase angle $\theta_0$ remains unspecified.

 We begin our investigation by setting $\psi$ uniform but at some stage we will
also ignore the uniform fluctuation $\delta \psi$ and deal only
with the equilibrium value $\psi_0$ of $\psi$. The equilibrium
value will be calculated after taking into account magnetic
fluctuations, so it will be different from the usual result
$|\psi_0| = (|a|/b)^{1/2}$~\cite{Lifshitz:1980}, where both
magnetic and superconducting fluctuations are ignored. This
simplest approximation for the equilibrium value of $\psi$ is
obtained from the GL free energy~(1), provided $e = 0$ and the
gradient term is neglected. Hereafter we will keep the symbol
$|\psi_0|$ for the equilibrium order parameter in the more general
case, when the magnetic fluctuations are not neglected, and will
denote the same quantity for $e=0$ by $\eta \equiv |\psi_0(e=0)| =
(|a|/b)^{1/2}$.

The above described approximation neglects the saddle point
solutions of GL equations, where $\langle \psi(\vec{x}) \rangle$
is $\vec{x}$-dependent. Therefore, the vortex state that is stable
in type II superconductors  cannot be achieved. This is consistent
with setting the external magnetic field to zero, so the vortex
state cannot occur in any type superconductor. Our arguments can
be easily verified with the help of GL
equations~\cite{Lifshitz:1980} for a zero external magnetic field;
the only nonzero solution for $\psi$ in this case is given by
$\eta = (|a|/b)^{1/2}$, although the magnetic fluctuations
$\vec{A}(\vec{x}) = \delta \vec{A}(\vec{x})$ are properly
considered.

In conclusion we can argue that the described method will be
convenient for both type I and type II superconductors in a zero
external magnetic field, if the $\psi$-fluctuations have a
negligibly small effect on phase transition properties
$T_{c0}=T_{c}(H_0=0)$, where  $T_{c}$ denotes the phase transition
line for any  $H_0 \ge 0$. For type II superconductors in  $H_0 >
0$, two lines  $T_{c1}(H_0)$ and $T_{c2}(H_0)$ should be defined,
usually given by $H_{c1}(T)$ and $H_{c2}(T)$~\cite{Lifshitz:1980}.

\subsection{Effective free energy}

 When the order parameter $\psi$ is uniform the
functional~(1)         is reduced to
\begin{equation}
\label{eq11} F(\psi, \vec{A}) =  F_0(\psi) + F_A(\psi)
\end{equation}
with
\begin{equation}
 \label{eq12}
  F_0(\psi) = V_D(a|\psi|^2 +\frac{b}{2}|\psi|^4)
 \end{equation}
and
\begin{equation}
\label{eq13} F_A(\psi) = \frac{1}{8\pi}\int d^Dx \left \{ \rho(\psi)
\vec{A}^2(\vec{x}) + \frac{1}{2} \sum_{i,j=1}^D \left( \frac{\partial
A_j}{\partial x_i} -\frac{\partial A_i}{\partial x_j} \right)^2
\right\} \;.
\end{equation}
Here $\rho = \rho_0 |\psi|^2$ and $\rho_0 = (8 \pi e^2/mc^2)$. It is
convenient to calculate the partition function $Z(\psi)$ and the
effective free energy ${\cal{F}}_{\mbox{\scriptsize D}}(\psi)$ in the
$\vec{k}$-space, where Eqs.~(\ref{eq9}) and~(\ref{eq13}) take the form
\begin{equation}
\label{eq14} \int_{- \infty}^{\infty} \prod_{j=1}^D
\prod_{\vec{k}>0}^{k \le \Lambda} d \mbox{Re} A_j(\vec{k}) d \mbox{Im}
A_j(\vec{k}) \delta \left[ \vec{k}\cdot \vec{A}(\vec{k}) \right ]
\end{equation}
and
\begin{equation}
\label{eq15} F_A(\psi) = F_A(0) + \Delta F_A(\psi)\;.
\end{equation}
Here
\begin{equation}
\label{eq16} F_A(0) = \frac{1}{8 \pi} \sum_{j,k} k^2 \left|A_j(\vec{k})
\right |^2 \;,
\end{equation}
and
\begin{equation}
\label{eq17} \Delta F_A(\psi)= \rho \sum_{j,k} \left|A_j(\vec{k})
\right |^2 \; ;
\end{equation}
note, that we have used the Coulomb gauge  $\vec{k}. \vec{A}(\vec{k}) =
0$.

Then the partition function~(\ref{eq8}) becomes
\begin{equation}
\label{eq18} {\cal{Z}}(\psi) = e^{-F_0(\psi)/k_B T}{\cal{Z}}_A(\psi)\;,
\end{equation}
where
\begin{equation}
\label{eq19} {\cal{Z}}_A(\psi) = \int {\cal{D}}A e^{-F_A(\psi)/k_B T}
\end{equation}
with $F_A(\psi)$ given by~(\ref{eq15}) and the functional
integration defined by the rule~(\ref{eq14}). With the help of
Eqs.~(\ref{eq10}) - (\ref{eq19}) the effective free energy
${\cal{F}}_D(\psi)$ will be
\begin{equation}
\label{eq20} {\cal{F}}_D(\psi) = F_0(\psi) + {\cal{F}}_f(\psi)\;,
\end{equation}
where $F_0(\psi)$ is given by Eq.~(\ref{eq12}) and
\begin{equation}
\label{eq21} {\cal{F}}_f(\psi) = -k_BT \ln{\left[
\frac{{\cal{Z}}(\psi)}{{\cal{Z}}(0)} \right]}
\end{equation}
is the $\psi$-dependent fluctuation part of  ${\cal{F}}(\psi)$. In
Eq.~(\ref{eq20}) the  $\psi$-independent fluctuation energy
$\{-k_B T \ln{\left[ {\cal{Z}}_A(0)\right]}\} $ has been omitted.
This energy should be ascribed to the normal state of the
superconductor, which by convention is set equal to zero.

Defining the statistical averages as
\begin{equation}
\label{eq22} \langle(...)\rangle = \frac{\int
{\cal{D}}{\cal{A}}e^{-F_A(0)/k_BT}(...)}{{\cal{Z}}_A(0)},
\end{equation}
we can write Eq.~(\ref{eq21}) in the form
\begin{equation}
 \label{eq23}
{\cal{F}}_f(\psi) = - k_BT \ln{\langle e^{- \Delta F_A(\psi)/k_BT }
\rangle}.
\end{equation}

Eq.~(\ref{eq23}) is a good starting point for the perturbation
calculation of ${\cal{F}}_f(\psi)$. We expand the exponent in
Eq.~(\ref{eq23}) and also take into account the effect of the
logarithm on the infinite series. In result we obtain
\begin{equation}
 \label{eq24}
{\cal{F}}_f(\psi) = \sum_{l=1}^{\infty} \frac{(-1)^l}{l!
(k_BT)^{l-1}}\langle \Delta F_A^l(\psi) \rangle_c \;,
 \end{equation}
where $\langle...\rangle_c$ denotes connected
averages~\cite{Uzunov:1993}. Now we have to calculate averages of
type
\begin{equation}
 \label{eq25}
\langle A_{\alpha}(\vec{k}_1),A_{\beta}(\vec{k}_2)...
A_{\gamma}(\vec{k}_n)\rangle_c \;.
\end{equation}
Here we will use the Wick theorem and the correlation function of
form
\begin{equation}
 \label{eq26}
G_{ij}^{(A)}(\vec{k}, \vec{k^{\prime}}) = \langle
A_i(\vec{k})A_j(-\vec{k^{\prime}}) \rangle =
\delta_{\vec{k},\vec{k^{\prime}}}G_{ij}^{A}(k)\;,
\end{equation}
where
\begin{equation}
 \label{eq27}
G_{ij}^{A}(\vec{k}) = \langle A_i(\vec{k})A_j(-\vec{k}) \rangle = \frac{4 \pi k_BT}{k^2}
\left( \delta_{ij} - \hat{k_i}  \hat{k_j}\right)
\end{equation}
 and $
\hat{k_i} = (k_i/k)$.

The calculation of the lowest order terms $(l=1,2,3)$ in
Eq.~(\ref{eq24}) with the help of ~(\ref{eq25})~-~(\ref{eq27}) is
straightforward. The perturbation terms in (\ref{eq24}) are shown
by diagrams in Fig.~1. The infinite series~(\ref{eq24}) can be
exactly summed up and the result is the following logarithmic
function
\begin{equation}
 \label{eq28}
{\cal{F}}_f(\psi) = \frac{(D-1)}{2}\: k_BT \sum_ k \ln
\left[1+\frac{\rho(\psi)}{k^2} \right]\;.
\end{equation}

\begin{figure}
\begin{center}
\epsfig{file=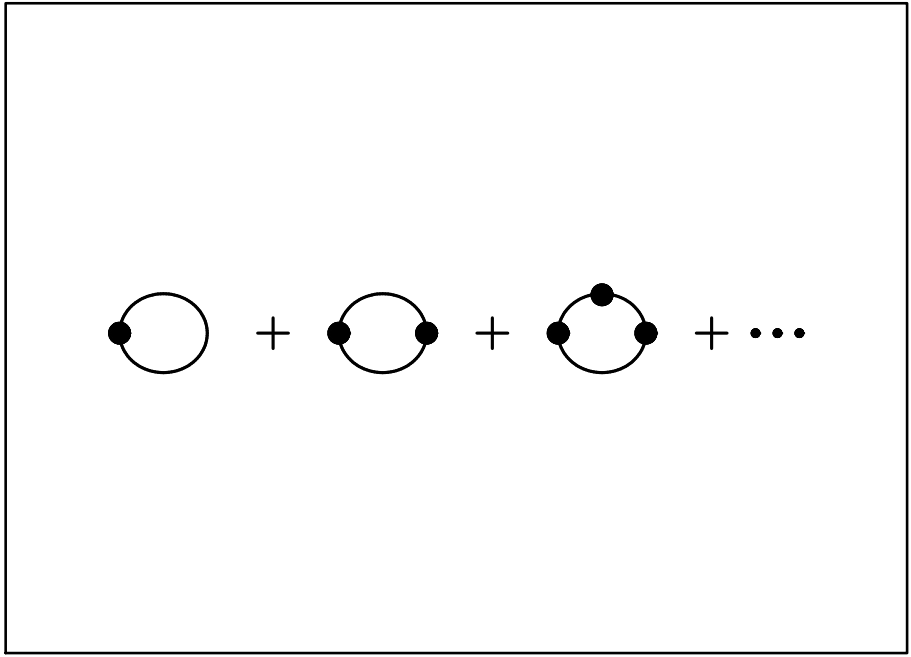,angle=-90, width=12cm}\\
\end{center}
\caption{Diagrammatic representation of the series (24); $\bullet$
represents the $\rho$-vertex in (13) and (17), and the solid lines
represent bare correlation functions $\langle
|A_j(\vec{k})|^2\rangle$.}
\end{figure}

The same result for ${\cal{F}}_f(\psi)$ can be obtained by direct
calculation of the Gaussian functional integral~(\ref{eq8}). This
is done using the integral representation of $\delta$-function
in~(\ref{eq9}) or~(\ref{eq14}) but it introduces an additional
functional integration that should be carried out after the
integration over $A_j(\vec{x})$.

Eqs.~(\ref{eq10}),~(\ref{eq20}) and~(\ref{eq28}) give the
 effective free energy density
\begin{equation}
 \label{eq29}
f_D(\psi) = {\cal{F}}_D(\psi)/V_D
\end{equation}
in the form
\begin{equation}
 \label{eq30}
f_D(\psi) = f_0(\psi) + \Delta f_D(\psi)\;,
\end{equation}
where
\begin{equation}
 \label{eq31}
 f_0(\psi) = a|\psi|^2 + \frac{b}{2} |\psi|^4
\end{equation}
and
 \begin{equation}
 \label{eq32}
\Delta f_D(\psi) =  \frac{(D-1) k_BT}{2 V_D}\sum_ k \ln
\left(1+\frac{\rho}{k^2} \right)\;.
\end{equation}

Eqs.~(\ref{eq20}) and ~(\ref{eq29})~-~(\ref{eq32}) are the basis
of our further considerations. We should mention that the
fluctuation contribution $\Delta f_D(\psi)$  to $f(\psi)$
transforms to convergent integral in the continuum limit
\begin{equation}
\label{eq33} \frac{1}{V_D} \sum_k \to \int \frac{d^Dk}{(2 \pi)^D} = K_D
\int_0^{\Lambda} dk.k^{D-1} \;,
\end{equation}
where $K_D = 2^{1-D} \pi^{-D/2}/ \Gamma(D/2)$ for all spatial
dimensionalities $D \ge 2$. But the terms in the expansion of the
logarithm in~(\ref{eq32}) are power-type divergent with the
exception of several low-order terms in certain dimensionalities
$D$. Therefore, we will work with a finite sum of an infinite
series with infinite terms. In our further calculations we will
keep the cutoff $\Lambda$ finite for all relevant terms in $\Delta
f_D(\psi)$. This is the condition to obtain correct results.

\subsection{2D-3D crossover}

The dimensional (2D-3D) crossover has been considered in
Refs.~\cite{Shopova4:2003, Rahola:2001}. Here we follow
Ref.~\cite{Shopova4:2003}, where the effective free energy density
$f(\psi)\equiv f_3(\psi) = {\cal{F}}_3(\psi)/V_3$ of a thin
superconducting film of thickness $L_0$ and volume $V =V_3 = (L_0
L_1L_2)$  is derived in a more general way, which allows the
investigation of the dimensional crossover. Now one should perform
the integration (33) only with respect to the wave vector
components $k_1$ and $k_2$, corresponding to the large sizes $L_j
\gg L_0$ ($j=1,2$). The result for the effective free energy
density is ~\cite{Shopova4:2003}:
\begin{equation}
\label{eq34}
 f(\psi) = a|\psi|^2 + \frac{b}{2}|\psi|^4 +
k_BTJ\left[\rho\left(\psi\right)\right]\;,
\end{equation}
where
\begin{equation}
\label{eq35}
 J(\rho) = \int_0^{\Lambda}\frac{dq}{2\pi}q
  S\left(q,\rho \right)
\end{equation}
is given by the sum
\begin{equation}
\label{eq36}
 S = \frac{1}{L_0} \sum_{k_0 = -\Lambda_0}^{+
 \Lambda_0}\mbox{ln}\left[1 + \frac{\rho(\psi)}
 {q^2 +k^2_0}\right],
\end{equation}
and $q = |\vec{q}|$, $\vec{q} = (k_1,k_2)$.

In Eqs.~(\ref{eq34})--(\ref{eq36}), the integral $J(\rho)$ and the
sum $S(q,\rho)$ over the wave vector $\vec{k} = (\vec{q},k_0)$ are
truncated by the upper cutoffs $\Lambda$ and $\Lambda_0$. The
finite cutoff $\Lambda$ is introduced for the wave number $q$ and
$\Lambda_0$ stands for $k_0$.

As our study is based on the quasimacroscopic GL approach  the
second cutoff $\Lambda_0$ should be again related to $\xi_0$
rather than to the lattice constant $a_0$, i.e. $\Lambda_0 \sim
(1/\xi_0)$, which means that phenomena at distances shorter than
$\xi_0$ are excluded from our consideration. We will assume that
the lowest possible value of $\Lambda_0$ is $(\pi/\xi_0)$, as is
for $\Lambda$, but we will keep in mind that both $\Lambda_0$ and
$\Lambda$  can be extended to infinity, provided the main
contribution to the integral $J(\rho)$ and the sum $S$ comes from
the long wavelength limit $(q\xi_0\ll 1)$.

In a close vicinity of the phase transition point $T_{c0}$ from
normal ($\psi = 0$) to Meissner state ($|\psi|
> 0$) the parameter $\rho \sim |\psi|^2$ is small and the main
contribution to the free energy $f(\psi)$ will be given by the
terms in $S$ with small wave vectors $ k \ll \Lambda$. This allows
an approximate but reliable treatment of the 2D-3D crossover by
expanding the summation over $k_0$ in~(\ref{eq36}) to infinity  -
$\Lambda_0 \sim \infty$. A variant of the theory when $\Lambda_0$
is kept finite ($\Lambda = \Lambda_0 = \pi/\xi_0$) can also be
developed but the results are too complicated~\cite{Todorov:2003}.
Performing the summation and the integration in
Eqs.~(\ref{eq35})--(\ref{eq36}) we obtain $J(\rho) =
(\Lambda^2/2\pi L_0)I(\rho)$, where
\begin{equation}
\label{eq37}
 I(\rho) = \int_0^{1}dy\:
 \mbox{ln}\left[\frac{\mbox
 {sh}\left(\frac{1}{2}\; L_0\Lambda\sqrt{\rho + y}\right)}{\mbox{sh}
 \left(\frac{1}{2}\; L_0\Lambda \sqrt{y}\right)}
  \right]\;,
\end{equation}
The integral~(\ref{eq37}) has a logarithmic divergence that
corresponds to the infinite contribution of  magnetic fluctuations
to the free energy of  normal phase ($ T_{c0} > 0, \varphi = 0$).
Such  type of divergence is a common property of many phase
transition models. In the present case, as is in other systems,
this divergence is irrelevant, because the divergent term does not
depend on the order parameter $\psi$ and the free energy $f(\psi)$
is defined as the difference between the total free energies of
the superconducting and normal phases: $f(\psi) = (f_S - f_N)$.

Introducing a dimensionless order parameter $\varphi =
(\psi/\psi_0)$, where $\psi_0 = (\alpha_0T_{c0}/b)^{1/2}$ is the
value of $\psi$ at $T=0$, we obtain the free energy~(\ref{eq34})
in the form
 \begin{equation}
\label{eq38}
 f(\varphi) = \frac{H_{c0}^2}{8\pi}\left[2t_0\varphi^2 + \frac{b}{2}|\varphi|^4 +
2(1+t_0)CI(\mu\varphi^2)\right]\;,
\end{equation}
with $I(\mu \varphi^2)$, given by Eq.~(\ref{eq37}), $\mu = (1/\pi
\kappa)^2$, $\Lambda = \pi/\xi_0$, and
\begin{equation}
\label{eq39}
 C = \frac{2\pi^2 k_BT_{c0}}{L_0\xi_0^2H^2_{c0}}\:.
\end{equation}
From the equation of state ($\partial f/\partial \varphi = 0$) we
find two possible phases: $\varphi_{00} = 0$ and the
superconducting phase $(\varphi_0 > 0)$, defined by the equation
\begin{equation}
 \label{eq40}
 t_0 + \varphi_0^2 + \frac{(1+t_0)CL_0\xi_0}{4\pi \lambda_0^2}K(\mu\varphi_0^2) =
 0 \:,
\end{equation}
where
\begin{equation}
\label{eq41}
 K(z) = \int_0^{1}dy\:\frac{\mbox
 {coth}\left(\frac{1}{2}\;L_0\Lambda\sqrt{y + z}\right)}
 {\sqrt{y + z}}\:.
\end{equation}
The analysis of the stability condition
($\partial^2f/\partial\varphi^2 \ge 0$) shows that the normal
phase is a minimum of $f(\varphi)$ for $t_0 \geq 0$, whereas the
superconducting phase is a minimum of $f(\varphi)$, if
\begin{equation}
\label{eq42}
 1 > \frac{1}{4}(1+t_0)CL_0\Lambda\mu^2\tilde{K}(\mu \varphi_0^2)\:,
 \end{equation}
 where
 \begin{equation}
 \label{eq43}
 \tilde{K}(z) = \int_0^{1}\frac{dy}{y + z}\left[\frac{\mbox
 {coth}\left(\frac{1}{2}\;L_0\Lambda\sqrt{y + z}\right)}
 {\sqrt{y + z}} + \frac{L_0\Lambda}{2\mbox{sh}^2\left(\frac{1}{2}\;L_0\Lambda\sqrt{y
 + z}\right)} \right]\:.
\end{equation}
  The entropy jump $\Delta S = (\Delta S/V) = [-df(\varphi_0)/dT]$ per unit volume at the
  equilibrium point of phase transition $T_c \neq T_{c0}$ is obtained in the form
\begin{equation}
\label{eq44}
 \Delta sS(T_c) = - \frac{H_{c0}^2\varphi_{c0}^2}{4\pi T_{c0}}
 \left[1+\frac{CI(\varphi_{c0})}
 {\varphi_{c0}^2}
 \right] \:,
\end{equation}
where $\varphi_{c0} \equiv \varphi_0(T_c)$ is the jump of the
dimensionless order parameter at $T_c$.

The second term in $\Delta S$ can be neglected. In fact, taking
into account the equation $f[\varphi_0(T_c)] = 0$ for the
equilibrium phase transition point $T_c$  we obtain that
$|CI(\varphi_0)/\varphi_0^2|$ is approximately equal to $|t_{c0} +
\varphi_{c0}^2/2|$, where $\varphi_{c0}^2$ and the dimensionless
shift of the transition temperature $t_{c0} = t_0(T_c)$ are
expected to be much smaller than unity. The latent heat $Q =
T_c\Delta S(T_c)$ and the jump of the specific heat capacity at
$T_c$, $\Delta C = T_c(\partial \Delta S/\partial T)$ can be
easily calculated with the help of Eq.~(\ref{eq44}). For this
purpose we need the function $\varphi_0(T)$, which cannot be
obtained in analytical form from Eq.~(\ref{eq40}).

Eqs.~(38) and (40) can be analyzed numerically. These relatively
simple 2D-3D crossover formulae can be used in investigations of
specific substances by  variation of thickness $L_0$ of films from
$L_0 \gg \Lambda^{-1}$ ($3D$ system) to $ a_0 < L_0 \ll
\Lambda^{-1} $ (quasi-$2D$ system) and even to a 2D system for
$L_0 = a_0$. Then one can vary the effective dimension of the
system $D_{eff}(L_0\Lambda)$ as a function of $L_0 \Lambda_0$ from
$D=2$ to $D=3$~\cite{Craco:1999}. However, from a purely
calculational point of view we have found more convenient to
consider particular dimensions of interest separately and then to
compare the results in order to demonstrate the relevant
differences between the bulk ($3D$) and thin film properties. This
approach is applied below.

\subsection{Effective free energy for particular dimensions}

For purely 2D superconductor, consisting of a single atomic layer,
we can use Eqs.~(\ref{eq29})-~(\ref{eq32}) by setting $D=2$, and
calculate $\Delta f_2(\psi)$ with the help of rule~(\ref{eq33}):
\begin{equation}
\label{eq45} \Delta f_2(\psi) = \left(\frac{k_BT}{8 \pi}\right)
\left[(\Lambda^2+ \rho_0|\psi|^2) \ln{\left
(1+\frac{\rho_0|\psi|^2}{\Lambda^2}\right)}
 -  \rho_0|\psi|^2 \ln{\left (\frac{\rho_0|\psi|^2}{\Lambda^2}\right)}
\right]\;.
\end{equation}
The first term of the above free energy can be expanded in powers
of $|\psi|^2$:
\begin{equation}
\label{eq46} \Delta f_2(\psi) = \left(\frac{k_BT}{8
\pi}\right)\left\{\rho_0|\psi|^2+ \rho_0|\psi|^2 \ln{\left
(\frac{\Lambda^2}{\rho_0|\psi|^2}\right)}
 + \frac{\rho_0^2|\psi|^4}{2 \Lambda^2}\right \}\;.
\end{equation}

Thus we obtain the result from Ref.~\cite{Lovesey:1980}. This case
is of special interest because of the logarithmic term in the
Landau expansion for $f(\psi)$, but it has no practical
application for the lack of ordering in purely 2D superconductors.

For quasi-2D superconductors we assume that $(2\pi/\Lambda) > L_0
\gg a_0$, where $L_0$ is the thickness of the superconducting
film, and the more precise choice of the upper cutoff $\Lambda \ll
(1/a_0)$ for the wave numbers $k_i$ is a matter of  additional
investigation ~\cite{Folk:2001}. In order to justify this
definition of quasi-2D system one can use the 2D-3D crossover
description presented in Sec.~2.4. The summation over the wave
number $k_0= (2 \pi n_0/L_0)$ in Eq. (36) cannot be substituted
with an integration because $L_0 \ll L_{j}$ and the dimension
$L_0$ does not obey the conditions, valid for $L_{j}$, $(j
=1,2)$~\cite{Suzuki:1994,Suzuki:1995,Craco:1999}. Therefore, for
such 3D system we must sum over $k_0$ and integrate over two other
components ($k_1$ and $k_2$) of the wave vector $\vec{k}$ (see
Sec.~2.4). This gives an opportunity for a systematic description
of the 2D-3D crossover as shown in Sec.~2.4. In the limiting case
of very small thickness the 2D-3D crossover theory (Sec. 2.4)
leads to a result, which is obtained more simply in an alternative
way, namely, by ignoring all terms corresponding to $k_0 \neq 0$
in the sum in Eq.~(\ref{eq32}). This corresponds to the
supposition that the quasi-2D film thickness cannot exceed
$2\pi\Lambda$. Assuming this point of view the real physical size
of the quasi-2D film thickness will depend on the choice of the
cutoff $\Lambda$. It is certain at this stage  that $\Lambda \geq
\xi_0$, because the (quasi)phenomenological GL theory does not
account phenomena of size less than $\xi_0$. The upper cutoff
$\Lambda$ of wave numbers can be defined in a more precise way at
next stages of consideration; see Sec.~II.6 - II.8.

For a quasi-2D film we have the expression:
\begin{equation}
\label{eq47} \Delta f(\psi) = \frac{2}{L_0} \Delta f_2(\psi)\;,
\end{equation}
where  $\Delta f_2(\psi)$ is given by Eq.~(\ref{eq45}).

For bulk (3D) superconductor we obtain:
\begin{equation}
\label{eq48} \Delta f_3(\psi) = \frac{k_BT}{2 \pi}\left[
\frac{\Lambda^3}{3}\ln{\left(1 + \frac{\rho_0|\psi|^2}{\Lambda^2}
\right)} +  \frac{2}{3}\rho_0|\psi|^2 \Lambda -
\frac{2}{3}\rho_0^{3/2} |\psi|^3 \arctan{\left(
\frac{\Lambda}{\sqrt{\rho_0 |\psi|^2}}\right)} \right].
\end{equation}
The Landau expansion in powers of $|\psi|$  in this form of
$f_3(\psi)$ confirms the respective results of
Refs.~\cite{Halperin:1974,Chen:1978}, moreover it correctly gives
the term of type $\rho^2_0 |\psi|^4$, which has been considered
small and neglected in these papers.

For 4D-systems $\Delta f_{\mbox{\scriptsize D}}(\psi)$ becomes
\begin{equation}
\label{eq49} \Delta f_4(\psi) = \frac{3k_BT}{64
\pi^2}\left[\Lambda^2 \rho_0 |\psi|^2 + \Lambda^4 \ln{\left(1 +
\frac{\rho_0|\psi|^2}{\Lambda^2} \right)} - \rho_0^2|\psi|^4
\ln{\left(1 +
 \frac{\Lambda^2}{\rho_0|\psi|^2} \right)}
\right].
\end{equation}
The above expression for $ \Delta f_4(\psi)$ can be also expanded
in powers of $|\psi|$ to show that it contains a term of the type
$|\psi|^4 \ln{(\sqrt{\rho_0}|\psi|/\Lambda)}$, which produces a
first order phase transition; this case is considered in  the
scalar electrodynamics~\cite{Coleman:1973}, as mentioned in Sec.
I. In our further investigation we will focus our attention on 3D
and quasi-2D superconductors.

The free energy density $\Delta f_{\mbox{\scriptsize D}}(\psi)$
can be expanded in powers of $|\psi|$ but the Landau expansion can
be done only in an incomplete way for even spatial dimensions.
Thus $f_2(\psi)$, $f_4(\psi)$, and $f(\psi)$ - the free energy
density corresponding to the quasi-2D films, contain logarithmic
terms, which should be kept in their original form in the further
treatment of the function $\Delta f_{\mbox{\scriptsize D}}(\psi)$
in the Landau expansion. The analysis has been
performed~\cite{Shopova5:2003} in two ways: with and without
Landau expansion of $\Delta f_D (\psi)$. These variants of the
theory are called ``exact" theory (ET) and ``Landau" theory (LT),
respectively~\cite{Shopova3:2003}. It has been
shown~\cite{Shopova5:2003} that these two ways of investigation
give the same results in all cases except for quasi-2D films with
relatively small thicknesses ($L_0 \ll \xi_0$). It seems important
to establish the differences between two variants of the theory
because the HLM effect is very small and any incorrectness in the
theoretical analysis may be a cause for an incorrect result. By
same arguments one can investigate the effect of the factor $T$ in
$\Delta f_{\mbox{\scriptsize D}}(\psi)$ on the thermodynamics of
quasi-2D films. The factor $T$ can be represented as $T= T_{c0}(1
+ t_0)$ and one may expect that the usual approximation $T\approx
T_{c0}$, which is well justified in the Landau theory of phase
transitions~\cite{Lifshitz:1980,Uzunov:1993}, may be applied. This
way of approximation can be performed by neglecting terms in the
thermodynamic quantities smaller than the leading ones. On the
other hand practical calculations lead to the conclusion that this
approximation cannot be made without a preliminary examination
because for some quasi-2D films it produces a substantial error of
about 10$\%$~\cite{Shopova5:2003}. LT, in which the factor $T$ is
substituted by $T_{c0}$ is referred to as ``simplified Landau
expansion" - SLT. All three variants of the theory, ET, LT and SLT
have been investigated in Ref.~\cite{Shopova5:2003}.

\subsection{Limitations of the theory}

The general result~(\ref{eq29})~-~(\ref{eq32}) for the effective
free energy $f(\psi)$ has the same domain of
validity~\cite{Lifshitz:1980} as the GL free energy functional in
a zero external magnetic field. When we neglect a sub-nano
interval of temperatures near the phase transition point we can
use Eq.~(1), provided $|t_0|=|T - T_{c0}|/T_{c0} <1$, or in the
particular case of type I superconductors, $|t_0|<
\kappa^2$~\cite{Lifshitz:1980}. Note, that the latter inequality
does not appear in the general GL approach. It comes as a
condition for the consistency of our approach with the microscopic
BCS theory for type I superconductors~\cite{Lifshitz:1980}.

Taking the continuum limit we have to assume that all dimensions of the
body, including the thickness $L_0$, are much larger than the
characteristic lengths $\xi$ and $\lambda$. The exception of this rule
is when we consider thin films. Especially for thin films of type I
superconductors, where ($(2\pi/\Lambda)
> L_0 \gg a_0$),  we should have in mind that
$\xi(T) >\lambda(T)$, so the inequalities $\xi> \lambda
> \xi_0 > \lambda_0$ hold true in the domain of validity of the GL theory
 $|t_0| < \kappa^2 < 1$.
In Ref.~\cite{Folk:2001} a comprehensive choice of the cutoff
$\Lambda$ has been made, namely, $\Lambda = \xi_0$ (the problem
for the choice of cutoff $\Lambda$ is discussed also in Sec.~II.7
-- II.8).
 The respective conditions for quasi-2D films
of type II superconductors are much weaker and are reduced to the usual
requirements: $\kappa > 1/\sqrt{2}$, $|t_0| < 1$ and $(2\pi/\Lambda) >
L_0 \gg a_0$.

 If we do a Landau expansion of $f_D(\psi)$ in powers of
$|\psi|^2$, the condition $\rho \ll \Lambda^2$ should be
satisfied. In order to evaluate this condition we substitute
$|\psi|^2$ in $\rho = \rho_0|\psi|^2$ with $\eta^2 =|a|/b$, which
corresponds to $e=0$.
 As $\lambda^2(T)=1/\rho$, the condition for the validity of
the Landau expansion becomes $[\Lambda \lambda(T)]^2 \gg 1$, i.
e., $(\Lambda \lambda_0)^2 \gg |t_0|$. Choosing the general form
of $\Lambda_{\tau} = (\pi \tau/\xi_0)$, where $\tau$ describes the
deviation of $\Lambda_{\tau}$ from $\Lambda_1 \equiv \Lambda =
(\pi /\xi_0)$, we obtain $(\pi \tau \kappa)^2 \gg |t_0|$; $\kappa
= (\lambda_0 /\xi_0)$ is the GL parameter.

Thus we can conclude that in type II superconductors,
where~$\kappa = (\lambda_0/\xi_0)> 1/\sqrt{2}$,~the condition
$(\rho/\Lambda^2) \ll 1$~is satisfied very well for values of the
cutoff in the interval between $\Lambda = (\pi/\xi_0)$ and
$\Lambda = (\pi/\lambda_0)$, i.e., for $1<\tau<(1/\kappa)$. For
type I superconductors, where $\kappa < 1/\sqrt{2}$ the cutoff
value $\Lambda \sim (1/\xi_0)$ leads to the BCS condition ($|t_0|
< \kappa^2$) for the validity of the GL approach. Substantially
larger cutoffs ($\Lambda \gg \pi/\xi_0$), for example, $\Lambda
\sim (1/\lambda_0)$ for type I superconductors with $\kappa \ll 1$
lead to a contradiction between the BCS condition and the
requirement $\rho \ll \Lambda^2$.

 In our calculations we often
use another parameter $ \mu_{\tau} = (1/\pi \tau \kappa)^2$ and,
in particular,  $\mu \equiv \mu_1 = (1/\pi \kappa)^2$ and in terms
of $\mu$ the condition for the validity of $f_D(\psi)$ expansion
becomes $\mu |t_0| \ll 1$, or, more generally, $\mu_{\tau}|t_0|
\ll 1$. Choosing $\tau = 1/\pi$, we obtain the BCS criterion for
the validity of the GL free energy for type I
superconductors~\cite{Lifshitz:1980}. The choice $\tau =
(\xi_0/\pi\lambda_0)$ corresponds to the cutoff $\Lambda_{\tau} =
1/\lambda_0$. As we will see in Sec.~II.7 the thermodynamics near
the phase transition point in 3D systems has no substantial
dependence on the value of the cutoff $\Lambda_{\tau}$ but it
should be chosen in a way that is consistent with the
mean-field-like approximation. The cutoff problem for quasi-2D
films has been investigated in Ref.~\cite{Shopova5:2003}. It has
been shown that the choice $\Lambda \sim \pi/\xi_0$ is consistent
in the framework of  GL theory.

Alternatively, the inequality $(\rho/\Lambda^2) \ll 1$ can be
investigated with the help of the reduced order parameter
$\varphi$ defined by $\varphi = |\psi|/\eta_0$, where $\eta_0
\equiv \eta(T=0)= (\alpha_0T_{c0}/b)^{1/2}$ is the so-called
zero-temperature value of order parameter for the GL free energy
$f_0(\psi)$, given by Eq.~(\ref{eq31}). The reduced order
parameter $\varphi$ will be equal to $|t_0|$ for $t_0 <0$, if only
the magnetic fluctuations are ignored, i.e., when $|\psi| = \eta$.
Using the notation $\varphi$, we obtain the condition
$(\rho/\Lambda^2) \ll 1$ in the form $\mu_{\tau}\varphi^2 \ll 1$.
This condition seems to be more precise because it takes into
account the effect of magnetic fluctuations on the order parameter
$\psi$.

\subsection{Bulk superconductors}

\subsubsection{Thermodynamics}

The effective free energy  $f_3(\psi)$ of bulk (3D-)
superconductors is given by Eqs.~(\ref{eq29})~-~(\ref{eq31})
and~(\ref{eq48}). The analytical treatment of this free energy can
be done by Landau expansion in small
$(\sqrt{\rho_0}|\psi|/\Lambda)$. Up to order $|\psi|^6$ we obtain
\begin{equation}
\label{eq50} f_3(\psi) \approx a_3 |\psi|^2 + \frac{b_3}{2}
|\psi|^4 - q_3|\psi|^3 + \frac{c_3}{2} |\psi|^6 \;,
\end{equation}
where
\begin{equation}
\label{eq51}
 a_3 = a + \frac{k_BT \Lambda \rho_0}{2\pi^2}\;,
\end{equation}
\begin{equation}
\label{eq52} b_3 = b + \frac{k_BT  \rho_0^2}{2\pi^2 \Lambda}\;,
\end{equation}
\begin{equation}
\label{eq53} q_3 = \frac{k_BT  \rho_0^{3/2}}{6 \pi}\;,
\end{equation}
and
\begin{equation}
\label{eq54} c_3 = - \frac{k_BT  \rho_0^{3}}{6 \pi^2 \Lambda^3}\;.
\end{equation}
The cutoff $\Lambda$ in Eqs.~(\ref{eq51}) - (\ref{eq54}) is not
specified and can be written in the form $\Lambda_{\tau} =
(\pi\tau/\xi_0)$ as suggested in Sec.~2.6.

 It can be shown by both analytical and numerical
calculations~\cite{Shopova:2002} that $|\psi|^6$-term has no
substantial effect on the thermodynamics, described by the free
energy~(\ref{eq50}). That is why we ignore this term. The possible
phases $|\psi_0|$ are found as a solution of the equation of
state:
\begin{equation}
\label{eq55} \left[ \partial f(\psi) / \partial |\psi| \:\right
]_{\psi_0} = 0\;.
\end{equation}
There always exists a normal phase $|\psi_0| = 0$, which is a
minimum of  $f_3(\psi)$ for $a_3 >0$. The possible superconducting
phases are given by
\begin{equation}
\label{eq56} |\psi_0|_{\pm} = \frac{3q_3}{4b_3} \left( 1 \pm
\sqrt{1-\frac{16a_3b_3}{9q_3^2}}\right) \ge 0.
\end{equation}
Having in mind  the existence and stability conditions of
$|\psi_0|_{\pm}$-phases ~\cite{Uzunov:1993}, we obtain that the
$|\psi_0|_{+}$-phase exists for $(16a_3b_3) \le 9q_3^2$ and this
region of existence always corresponds to a minimum of
$f_3(\psi)$. The $|\psi_0|_{-}$-phase exists for $0 < a_3 <
9q_3^2/16b_3$ and this region of existence always corresponds to a
maximum of   $f_3(\psi)$, i.e., this phase is absolutely unstable.
For $a_3 =0,\; |\psi_0|_{-}=0$ and hence, it coincides with the
normal phase. For $9q_3^2 = 16a_3b_3$ we have $|\psi_0|_{+} =
|\psi_0|_{-}=3q_3/4b_3$  and $ f_3(|\psi_0|_{+})=
f_3(|\psi_0|_{-})= 27q_3^4/512b_3^3$. Furthermore
$f_3(|\psi_0|_{-}) > 0$ for all allowed values of $|\psi_0|_{-}
> 0$, whereas $f_3(|\psi_0|_{+}) < 0$ for
$a_3<q_3^2/2b_3$, and $ f_3(|\psi_0|_{+}) > 0$ for $(q_3^2/2b_3) <
a_3< 9q_3^2/16b_3$. The equilibrium temperature
$T_{\mbox{\scriptsize eq}}$ of the first order phase transition is
defined by the equation $f(|\psi_0|_{+})= 0$, which gives the
following result:
\begin{equation}
\label{eq57}
 2b_3(T_{\mbox{\scriptsize eq}})a_3(T_{\mbox{\scriptsize eq}}) =
  q_3^2(T_{\mbox{\scriptsize eq}})\;.
 \end{equation}
These results are confirmed by numerical calculations of the
effective free energy~(\ref{eq50})~\cite{Shopova:2002}.

The equilibrium entropy jump is $\Delta S = V \Delta s$ and
$\Delta s = - (df_3(|\psi|)/dT)$ can be calculated with the help
of Eq.~(\ref{eq50}) and the equation of state~(\ref{eq55}):
\begin{equation}
\label{eq58} \Delta s = - |\psi_0|^2 \Phi(|\psi_0|) \;,
\end{equation}
where $\Phi(|\psi_0|)$ is the following function:
\begin{equation}
\label{eq59} \Phi(y) = (\alpha_0+\frac{k_B \Lambda \rho_0}{2
\pi^2}) - \frac{\rho_0^{3/2}k_B}{6 \pi} y + \left(\frac{k_B
\rho_0^2}{4 \pi^2 \Lambda}\right)  y^2\;.
\end{equation}

The specific heat capacity per unit volume $\Delta C =
T(\partial\Delta s/ \partial T)$ is obtained from~(\ref{eq58}):
\begin{equation}
\label{eq60} \Delta C = - \left( \frac{T}{T_{c0}}\right )
\frac{\partial |\psi_0|^2}{ \partial t_0} \Phi(|\psi_0|) \;.
 \end{equation}
The quantities  $\Delta s(T)$ and $\Delta C(T)$ can be evaluated
at the equilibrium phase transition point $T_{\mbox{\scriptsize
eq}}$, which is found from Eq.~(\ref{eq57}):
\begin{equation}
\label{eq61} \frac{T_{\mbox{\scriptsize eq}}}{T_{c0}} \approx 1-
\frac{k_B \rho_0 \Lambda}{2 \pi^2 \alpha_0} + \frac{\left(
\rho_0^{3/2}k_B/6 \pi \right)^2 }{b+\left(\rho_0^{2}k_B/2 \pi^2
\Lambda \right)T_{c0}} \left( \frac{T_{c0}}{\alpha_0}\right)\;,
  \end{equation}
provided $|\Delta T_c| =|T_{c0}-T_{\mbox{\scriptsize eq}}| \ll
T_{c0}$. Further we will see  that the condition $|\Delta T_c| \ll
T_{c0}$ is valid in real substances. The second term in r.h.s. of
Eq.~(\ref{eq61}) is a typical negative fluctuation contribution,
whereas the positive third term in  r.h.s. of the same equality is
typical of first-order transitions~\cite{Uzunov:1993}.

To obtain the jumps  $\Delta s$ and $\Delta C$ at
$T_{\mbox{\scriptsize eq}}$ we have to put the solution
$|\psi_0|_{+}$, found from Eq.~(\ref{eq56}) in
Eqs.~(\ref{eq58})~-~(\ref{eq60}). The result will be:
\begin{equation}
\label{eq62} \Delta s = - \frac{q_{3c}^2}{b_{3c}^2}\left \{
\alpha_0 + \frac{k_B\rho_0 \Lambda}{2 \pi^2}- \left(
\frac{k_B\rho_0^{3/2} }{6 \pi}\right)^2 \frac{T_{\mbox{\scriptsize
eq}}}{b_{3c}^2} \right \}\;,
\end{equation}
 and
\begin{equation}
\label{eq63} \Delta C = \frac{4 \alpha_0}{b_{3c}}\left(\alpha_0
T_{c0} - \frac{q_{3c}^2b}{b_{3c}^2}\right ) \;,
 \end{equation}
 where $b_{3c}$ and $q_{3c}$ are the parameters $b_{3}$ and $q_{3}$  at
 $T=T_{\mbox{\scriptsize eq}}$. As $|\Delta T_c| =|T_{c0}-T_{\mbox{\scriptsize eq}}|
  \ll T_{c0}$ we can set
 $T_{\mbox{\scriptsize eq}} \approx T_{c0}$ in r.h.s. of
 Eqs.~(\ref{eq62})~and~(\ref{eq63}) and
  obtain $q_{3c} \equiv q_3(T = T_{\mbox{\scriptsize eq}})
 \approx q_3(T_{c0})$ and $b_{3c} \approx b_3(T_{c0})$. The latent heat
  $Q=- T_{\mbox{\scriptsize eq}} \Delta s$ per unit
volume of the first order phase transition at
$T_{\mbox{\scriptsize eq}}$ can be calculated from
 Eq.~(\ref{eq62}).

 Now we shall discuss the ratio
 \begin{equation}
 \label{eq64}
 (\Delta T)_{\mbox{\scriptsize eq}} = \frac{Q}{\Delta C}
 \end{equation}
 introduced in Ref. \cite{Halperin:1974} as a measure of the
 temperature size of the HLM effect -- a temperature interval, where the effect can be
 observed (in \cite{Halperin:1974} the same ratio is denoted by
 $\Delta T_1$). Keeping only the first terms in Eqs.~(\ref{eq62})~-~(\ref{eq63}),
 we obtain a result for $(\Delta T)_{\mbox{\scriptsize eq}}$
 which is four times smaller than that for $\Delta T_1$~\cite{Halperin:1974}.
 To explain the difference let us mention that Eq.~(\ref{eq63}) gives the jump $\Delta C$
 at the equilibrium phase transition point of the first order phase
 transition, described by $|\psi|^3$ term~\cite{Uzunov:1993}, while
 the specific heat jump considered in~Ref.~\cite{Halperin:1974} is equal to the specific heat
 jump at the standard second order transition
 $\Delta \tilde{C} =(\alpha_0^2T_{c0}/b)$ \cite{Uzunov:1993},
 and is four times smaller. In fact, neglecting the second term in Eq. (63) and
  having in mind that $b_3\equiv b$,
  see Eq. (66) below, we have
 that $\Delta C = 4\Delta {\tilde{C}}$.  Therefore, we obtain that
 $(\Delta T)_{\mbox{\scriptsize eq}}$ is four
 times smaller than the respective value $\Delta T_1$ in
 Ref.~\cite{Halperin:1974}. This is valid in an approximation in which the second
 term in Eq. (63) is neglected (this approximation is justified in Sec.
 2.7.2). The approximation of $\Delta s$ and $\Delta C$ by the
 first (leading) terms in Eqs. (62) and (63) does not change
 essentially these quantities provided the cutoff $\Lambda$ is
 comprehensively chosen (see also Sec. 2.7.2). Note that in Refs.
 \cite{Halperin:1974, Chen:1978} as well as in all preceding
 papers, the corrections to the leading terms in $\Delta s$ and
 $\Delta C$ have not been taken into account.

 \subsubsection{Results for Al}

 In order to do the numerical estimates we represent the Landau
parameters $\alpha_0$ and $b$  with the help of the
zero-temperature coherence length $\xi_0$ and the zero-temperature
critical magnetic field $H_{c0}$. The connection between them is
given by formulae of the standard GL superconductivity theory
~\cite{Lifshitz:1980}: $\xi_0^2=(\hbar^2/4m\alpha_0 T_{c0})$ and
$H_{c0}^2=(4 \pi \alpha_0^2 T_{c0}^2/b)$. The expression for the
zero-temperature penetration depth $\lambda_0=(\hbar c/2 \sqrt{2}
e H_{c0}\xi_0 )$ is obtained from the above relation and
$\lambda_0=(b/\alpha_0 T_{c0} \rho_0)^{1/2}$. We will use the
experimental values of  $T_{c0}$, $H_{c0}$ and $\xi_0 $ for Al as
given in Table 1. The experimental values for $T_{c0}$, $H_{c0}$
and $\xi_0$ vary about 10-15$\%$, depending on the method of
measurement and the geometry of the samples (bulk material or
films)~\cite{Madelung:1990} but such deviations do not affect the
results of our numerical investigations.

 \small
 {\bf Table 1.} Values of $T_{c0}$, $H_{c0}$, $\xi_0$,
$\kappa$, and $|\psi_0|$ for W, Al, In~\cite{Madelung:1990}.\\
\begin{tabular}{cccccc}
\hline Substance & $T_{c0}$ (K) & $H_{c0}$ (Oe)& $\xi_0$ ($\mu$m)
& $\kappa$ & $|\psi_0|\times 10^{-11}$  \\ \hline W & $0.015$ &
$1.15$
& $37$ & $0.001 $& $0.69$  \\
\hline Al & $1.19$ & $99.00$ & 1.16 & $0.010$ &
$2.55$ \\
\hline In & 3.40 & $281.5$ & 0.44 & 0.145 & $2.0$
\\
\hline
\end{tabular}

\normalsize

\vspace{0.5cm}

The evaluation of the parameters $a_3$ and $b_3$ for Al
gives:
\begin{equation}
\label{eq65} a_3 = (\alpha_0 T_{c0})\left[t_0 + 0.972 \times
10^{-4} (1+t_0)\tau\right],
\end{equation}
and
\begin{equation}
\label{eq66} \frac{b_3}{b} =1 +\frac{0.117}{\tau}\;.
\end{equation}
Setting $\tau=1$ corresponds to the cutoff $\Lambda_1
=(\pi/\xi_0)$ (Sec~2.6). For $\tau =(1/\kappa)_{Al} = 10^2$, which
corresponds to the much higher cutoff $\Lambda =(\pi/\lambda_0)$,
we have $b_3 \approx b$, i.e., the~$\rho_0^2$~-term in $b_3$,
given by Eq.~(\ref{eq52}), can be neglected. However, as we can
see from Eq.~(\ref{eq66}), for $\tau = 1$ the same
$\rho_0^2$-correction in the parameter $b_3$ is of order $0.1b$
and cannot be automatically ignored in all calculations, in
contrast to the supposition in Refs.~\cite{Halperin:1974,
Chen:1978}. However, the more important fluctuation contribution
in 3D superconductors comes from the $\tau-$term in
Eq.~(\ref{eq65}) for the parameter $a_3$. This term is of order
$10^{-4}$ for $\tau \sim 1$ and this is consistent with the
condition $|t_0| < \kappa^2 \sim 10^{-4}$, but for $\tau \sim
10^2$, i.e., for $\Lambda \sim (\pi/\lambda_0) \sim 10^6\mu$m, the
same $\tau-$ term is of order $10^2$, which exceeds the
temperature interval ($T_{c0} \pm 10^{-4}$) for the validity of
BCS condition for Al (Sec.~2.6).

These results demonstrate that for our theory to be consistent, we
must choose the cutoff $\Lambda_{\tau}=(\pi \tau/\xi_0)$, where
$\tau$ is not a large number $(\tau \to 1 \div 10)$. To be more
specific, we set $\Lambda=\Lambda_1=(\pi/\xi_0)$ as suggested in
Ref.~\cite{Folk:2001}.

The temperature shift $t_{\mbox{\scriptsize eq}} =
t_0(T_{\mbox{\scriptsize eq}})$ for bulk Al can be estimated with
the help of Eq.~(\ref{eq61}). We obtain that this shift is
negative and very small: $t_{\mbox{\scriptsize eq}} \sim -
10^{-4}$. Note, that the second term in the r.h.s. of
Eq.~(\ref{eq61}) is of order $10^{-4}$, provided $\Lambda \sim
(1/\xi_0)$ whereas the third term in the r.h.s. of the same
equality is of order $10^{-5}$. Once again the change of the
cutoff $\Lambda$ to values much higher than ($\pi/\xi_0$) will
take the system outside the temperature interval, where the BCS
condition for Al is valid. Let us note, that in
Ref.~\cite{Shopova:2002} the parameter $t$ corresponds to our
present notation $t_0$. But the numerical calculation of the free
energy function $f_3(\psi)$ in Ref.~\cite{Shopova:2002} was made
for the SLT variant of the theory and the shifted parameter ($t_0
+ 0.972\times 10^{-4}$) was incorrectly identified with  $t$ and
this has led to the wrong conclusion for its positiveness at the
equilibrium phase transition point $T_{\mbox{\scriptsize eq}}$. As
a matter of fact, the shifted parameter ($t_0 + 0.972\times
10^{-4}$) is positive at $T_{\mbox{\scriptsize eq}}$ but
$t_{\mbox{\scriptsize eq}} \equiv t_0(T_{\mbox{\scriptsize eq}})$
is negative, as firstly noted in Ref.~\cite{Shopova5:2003}.

Having in mind these remarks, when we evaluate $\Delta  s$ and
$\Delta C$ for bulk Al we can use simplified versions of
~(\ref{eq62}) and ~(\ref{eq63}) which means to consider only the
first terms in the r.h.s and  to take $q_{3c} \approx q_3$   and
$b_{3c} \approx b$ at $T_{c0}$. In this way we obtain
\begin{equation}
\label{eq67} Q = - T_{c0}\Delta s  = 0.8 \times 10^{-2}
\left[\frac{\mbox{erg}}{\mbox{K}\:.\: \mbox{cm}^3}\right]\;,
\end{equation}
and
\begin{equation}
\label{eq68} \Delta C  = 2.62 \times 10^{3}
\left[\frac{\mbox{erg}}{\mbox{cm}^3}\right]\;.
\end{equation}
The results are consistent with an evaluation of $\Delta C$ for Al
as a jump ($\Delta \tilde{C} = \alpha_0^2T_{c0}/b$) at the second
order superconducting transition point~\cite{Halperin:1974} that,
as we have mentioned above, is four times smaller than the jump
$\Delta C$ given by Eq.~(\ref{eq68}).

A complete numerical evaluation of the function $f_3(\psi)$ and
the jump of the order parameter at $T_{\mbox{\scriptsize eq}}$ for
bulk Al was presented for the first time in
Ref.~\cite{Shopova:2002}. The results there confirm that the
order parameter jump and Q for bulk type I superconductors are
very small and can hardly be observed in experiments.

We will finish the presentation of bulk Al with a discussion of
the ratio~(\ref{eq64}). It can be also written in the form
\begin{equation}
\label{eq69}
 (\Delta T)_{\mbox{\scriptsize eq}} = \frac{32\pi}{9}  \left(\frac{k_B^2T_{c0}^2}{
b \alpha_0}\right) \left(\frac{e^2}{ m c^2}\right)^3,
\end{equation}
and it differs by a factor $1/4$ from the respective result in
Ref.~\cite{Halperin:1974}. The reason for this difference was
explained at the end of Sec. 2.7.1. From Eq.~(\ref{eq69}) we have
\begin{equation}
\label{eq70} (\Delta T)_{\mbox{\scriptsize eq}} = 6.7 \times
10^{-12} (T_c^3 H_{c0}^2 \xi_0^6),
\end{equation}
and multiplying the number coefficient in the above expression by
four we obtain Eq.~(10) presented in Ref.~\cite{Halperin:1974}.

\subsection{Quasi-2D films}

Following Refs.~\cite{Shopova2:2003, Shopova3:2003} we can present
the free energy density $f(\psi)=(F(\psi)/L_1L_2)$ in the form
\begin{equation}\label{eq71}
 f(\varphi) =  \frac{H_{c0}^2}{8 \pi}\left[2 t_0 \varphi^2 + \varphi^4
 + C(1+ t_0 ) \Gamma(\mu \varphi^2) \right],
\end{equation}
where
\begin{equation}
\label{eq72}
  \Gamma(y) = (1+y) \ln{(1+y)} -y \ln{y},
\end{equation}

To obtain Eqs.~(\ref{eq71})~-~(\ref{eq72}) we have set
$\Lambda=(\pi/\xi_0)$ and introduced the notation $\varphi =
|\psi|/\eta_0$; $\eta_0$ is defined in Sec.~2.5. Some of the
properties of free energy~(\ref{eq71}) were analyzed in
Ref.~\cite{Shopova2:2003} for Al films and in
Ref.~\cite{Shopova3:2003} for films of Tungsten (W), Indium (In),
and Aluminium (Al). Here we will briefly discuss the main results.

The equilibrium order parameter $\varphi_0 > 0$ corresponding to
the Meissner phase, can be easily obtained from the equation
$\partial f(\varphi)/\partial\varphi = 0$ and Eq.~(\ref{eq71}):
\begin{equation}
\label{eq73}
 t_0 + \varphi^2_0 + \frac{C\mu(1+t_0)}{2}\:\mbox{ln}\left(1 +
 \frac{1}{\mu\varphi_0^2}\right)
 = 0\:.
\end{equation}

 The logarithmic divergence in Eq.~(\ref{eq73}) has no chance to occur because $\varphi_0$ is
  always positive and does not tend to zero.

The largest terms in the entropy jump $\delta s$ and the specific
heat jump $\delta C$ at the equilibrium first order phase
transition point $T_{\mbox{\scriptsize eq}}$ are given
by~\cite{Shopova2:2003, Shopova3:2003}
\begin{equation}
\label{eq74}
 \delta s = - \frac{H_{c0}^2}{4\pi T_{c0}} \varphi^2_{\mbox{\scriptsize
eq}},
\end{equation}
 where $\varphi_{\mbox{\scriptsize eq}} = \varphi_0(T=T_{\mbox{\scriptsize
eq}})$, and
\begin{equation}
\label{eq75}
 \delta C = \frac{H^2_{c0}}{4\pi T_{c0}} \:.
\end{equation}
The latent heat of the phase transition~\cite{Uzunov:1993} is
given by $Q = - T_{\mbox{\scriptsize eq}}\delta s$ and
Eqs.~(\ref{eq73})-(\ref{eq74}). Since
 the temperatures $T_{\mbox{\scriptsize eq}}$ and $T_{c0}$ have
very close values, the difference between the values of $Q$,
$\delta s$, and $\delta C$ at $T_{c0}$ and $T_{\mbox{\scriptsize
eq}}$, respectively, can also be ignored, for example, $|\delta
C(T_{\mbox{\scriptsize eq}})-\delta C(T_{c0})|/\delta C(T_{c0})
\ll 1$ and we can use either $\delta C(T_{c0})$ or $\delta
C(T_{\mbox{\scriptsize eq}})$ ~\cite{Shopova2:2003}.

The equations~(\ref{eq71}) - (\ref{eq73}) corresponding to
quasi-2D films are quite different from the respective equations
for bulk (3D-) superconductors but it is easily seen that the
relatively large value of the order parameter jump
$\varphi^2_{\mbox{\scriptsize}}$ in thin films again corresponds
to relatively small value of the GL parameter $\kappa$. That is
why we consider element superconductors with small values of
$\kappa$ and study the effect of this parameter, the critical
magnetic field $H_{c0}$ and the film thickness $L_0$ on the
properties of the fluctuation-induced first order phase
transition.

We use experimental data for $T_{c0}$, $H_{c0}$, $\xi_0$ and
$\kappa$ for W, Al, and In, published in Ref.~\cite{Madelung:1990}
(see Table 1). In some cases the GL parameter $\kappa$ can be
calculated with the help of the relation  $\kappa =
(\lambda_0/\xi_0)$ and the available data for $\xi_0$ and
$\lambda_0$. In other cases it is more convenient to use the
following representation of the zero-temperature penetration
depth:
\begin{equation}
\label{eq76}
  \lambda_0 = \frac{\hbar c}{2\sqrt{2}eH_{c0}\xi_0}\:.
\end{equation}
The value of $|\psi_0|$ in Table~1 is found from
 \begin{equation}
\label{eq77}
  |\psi_0| = \left(\frac{m}{\pi \hbar^2}\right)^{1/2}\xi_0H_{c0}\:.
\end{equation}

The order parameter dependence on the reduced temperature
difference $t_{0}$ is shown in Fig.~2 for Al films of different
thicknesses. It is readily seen that the behavior of the function
$\varphi_0(t_0)$ corresponds to a well established phase
transition of first order. The vertical dashed lines in Fig.~2
indicate the respective values of $t_{\mbox{\scriptsize eq}} =
t_0(T_{\mbox{\scriptsize eq}})$, at which the equilibrium phase
transition occurs, as well as the equilibrium jump
$\varphi_0(T_{\mbox{\scriptsize eq}})= \varphi_{\mbox{\scriptsize
eq}}$ for different thicknesses of the film. The parts of the
$\varphi_0(t_0)$-curves, which extend up to $t_0 >
t_{\mbox{\scriptsize eq}}$, describe the metastable (overheated)
Meissner states, which can appear under certain experimental
circumstances (see in Fig.~2 the parts of the curves on the r.h.s.
of the dashed lines). The value of $\varphi_{\mbox{\scriptsize
eq}}$ and the metastable region decrease with the increase of the
film thickness, which shows that the first order of the phase
transition is better pronounced in thinner films and this confirms
a conclusion in Ref.~\cite{Shopova2:2003}.

\begin{figure}
\begin{center}
\epsfig{file=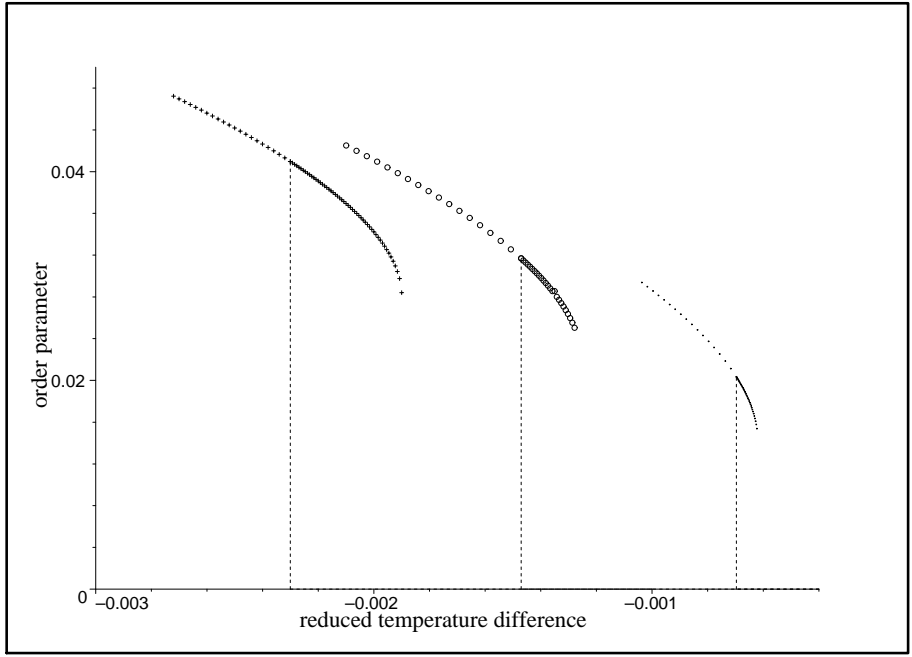,angle=-90, width=12cm}\\
\end{center}
\caption{Order parameter profile $\varphi(t_{0})$ of Al films of
different thicknesses: $L_0 = 0.05~\mu$m (``+"-line), $L_0 =
0.1~\mu$m ($\circ$), and $L_0 = 0.3~\mu$m
($\cdot$)~\cite{Shopova3:2003}.} \label{ST2f1.fig}
\end{figure}

\begin{figure}
\begin{center}
\epsfig{file=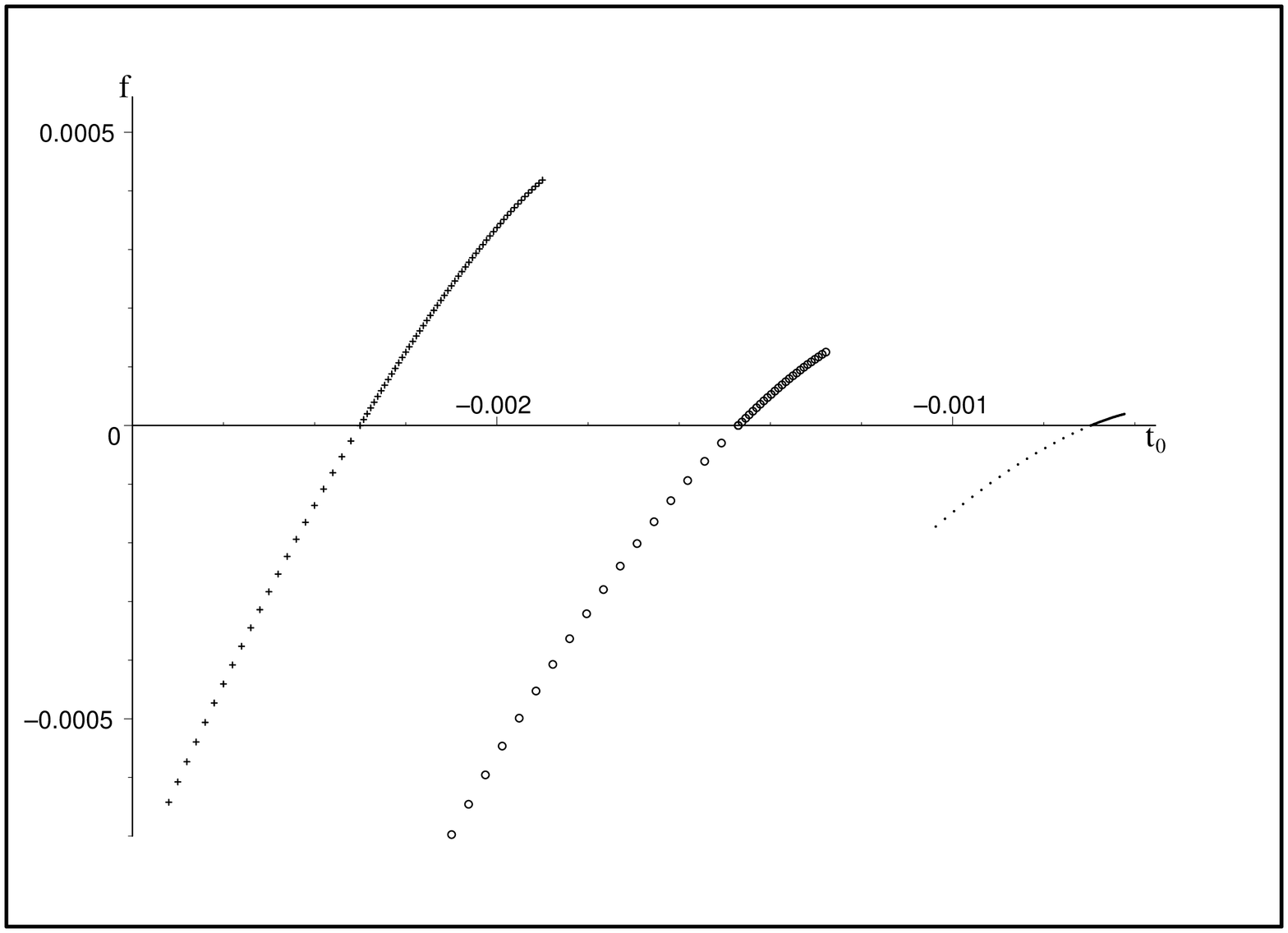,angle=-90, width=12cm}\\
\end{center}
\caption{The free energy $f(t_0)$ for Al films of thickness: $L_0
= 0.05~\mu$m (``+"-line), $L_0 = 0.1~\mu$m ($\circ$), $L_0 =
0.3~\mu$m ($\cdot$)~\cite{Shopova3:2003}.}
 \label{ST2f2.fig}
\end{figure}

\begin{figure}
\begin{center}
\epsfig{file=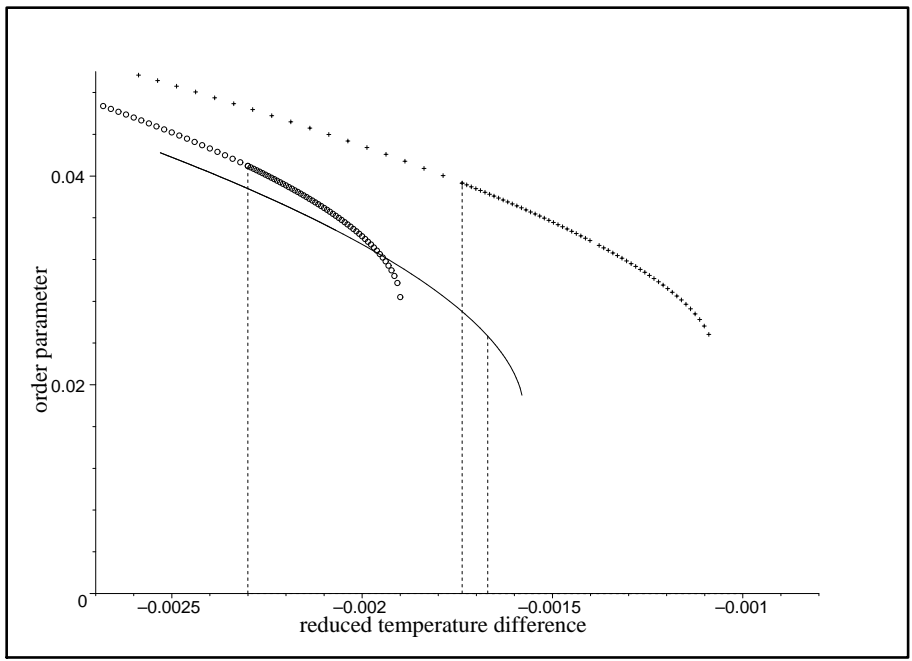,angle=-90, width=12cm}\\
\end{center}
\caption{Order parameter profile $\varphi(t_0)$ of films of
thickness $L_0 = 0.05~\mu$m: W (``+"-line), Al ($\circ$), and In
($\cdot$)~\cite{Shopova3:2003}.} \label{ST2f3.fig}
\end{figure}

These results are justified by the behavior of the free energy as
a function of $t_0$. We used Eqs.~(\ref{eq71})-(\ref{eq73}) for
the calculation of the equilibrium free energy
$f[\varphi_0(t_0)]$. The free energy for Al films with different
thicknesses is shown in Fig.~3. The equilibrium points
$T_{\mbox{\scriptsize eq}}$ of the phase transition correspond to
the intersection of the $f(\varphi_0)$-curves with the $t_0$-axis.
It is obvious from Fig.~3 that the temperature domain of
overheated Meissner states decreases with the increase of the
thickness $L_0$.

The shape of the equilibrium order parameter $\varphi_0(t_{0})$ in
a broad vicinity of the equilibrium phase transition for thin
films ($L_0 = 0.05\mu$m) of W, Al, and In was found from
Eq.~(\ref{eq73}). The result is shown in Fig.~4.  The vertical
dashed lines in Fig.~4 again indicate the respective values of
$t_{\mbox{\scriptsize eq}} = t_0(T_{\mbox{\scriptsize eq}})$, at
which the equilibrium phase transition occurs as well as the
equilibrium jump $\varphi_0(T_{\mbox{\scriptsize eq}})=
\varphi_{\mbox{\scriptsize eq}}$ in the different superconductors.

The order parameter jump at the phase transition point of In (the
In curve is marked by points in Fig.~4) is relatively smaller than
for W, and Al, where the GL parameter has much lower values. The
same is valid for the metastability domains; see the parts of the
curves in Fig.~4 on the left of the vertical dashed lines. It is
obvious from Fig.~4 and Table~2 that the equilibrium jump of the
reduced order parameter $\varphi_{\mbox{\scriptsize eq}}$ of W has
a slightly smaller value than that of Al, although the GL number
$\kappa$ for W has a ten times lower value, compared to $\kappa$
of Al. Note, that in Fig.~4 we show the jump of
$\varphi_{\mbox{\scriptsize eq}}$, but the important quantity is
$|\psi|_{\mbox{\scriptsize eq}}
=|\psi_0|\varphi_{\mbox{\scriptsize eq}}$. Using the data for $L_0
= 0.05\mu$m from Tables~1 and 2 we find for
$|\psi|_{\mbox{\scriptsize eq}}$ the following values: $0.1\times
10^{11}$ for Al, $0.05\times 10^{11}$ for In, and $0.02\times
10^{11}$ for W. This result shows that the value of the critical
filed $H_{c0}$ is also important and should be taken into account
together with the smallness of GL number when the maximal values
of the order parameter jump are looked for. Thus the value of the
order parameter jump at the fluctuation-induced phase transition
is maximal, provided small values of the GL parameter $\kappa$ are
combined with relatively large values of the critical field
$H_{c0}$. In our case Al has the optimal values of these two
parameters.

 \small
  { \small \bf Table 2.} \small Values of
$t_{\mbox{\scriptsize eq}}$, $\varphi_{\mbox{\scriptsize eq}}$,
  and $Q$ (erg/cm$^3$) for films of W, Al, and In
   with different\\ thicknesses $L_0$~($\mu$m)~\cite{Shopova3:2003}. \\
\small
\begin{tabular}{lccccccccc}
\hline
&\multicolumn{3}{c}{Al}&\multicolumn{3}{c}{In}&\multicolumn{3}{c}{W}
\\\cline{2-10}
 $L_0$ &$\mbox{t}_{\mbox{\scriptsize eq}}$
& $\mbox{$\varphi$}_{\mbox{\scriptsize eq}}$ & $Q$
&$\mbox{t}_{\mbox{\scriptsize eq}}$ & $\varphi_{\mbox{\scriptsize
eq}}$ & $Q$ &$\mbox{t}_{\mbox{\scriptsize eq}}$ &
$\varphi_{\mbox{\scriptsize eq}}$ & $Q$ \\
\hline
 0.05 &$-0.00230$ & $0.041$ & $1.95$ &$-0.00167$ &
$0.025$ & $3.94$ &$-0.00174$ & $0.039$ & $1.6\times 10^{-4}$
\\\hline 0.1 &$-0.00147$ & $0.032$ & $0.80$
&$-0.00094$ & $0.017$ & $1.82$ &$-0.00118$ & $0.032$ & $1.1\times
10^{-4}$
\\\hline 0.3&$-0.00070$ & $0.023$ & $0.41$
&$-0.00037$ & $0.010$ & $0.63$ &$-0.00064$ & $0.023$ & $5.6\times
10^{-5}$
\\\hline 0.5 &$-0.00048$ & $0.016$ & $0.20$
&$-0.00029$ & $0.008$ & $0.40$ &$-0.00048$ & $0.020$ & $4.1\times
10^{-5}$
\\\hline 1 &$-0.00029$ & $0.012$ & $0.11$
&$-0.00013$ & $0.006$ & $0.23$ &$-0.00032$ & $0.016$ & $2.7\times
10^{-5}$
\\\hline 2 &$-0.00017$ & $0.009$ & $0.06$
&$-0.00008$ & $0.004$ & $0.10$ &$-0.00021$ & $0.013$ & $1.8\times
10^{-5}$ \\
\hline
\end{tabular}

\vspace{0.5cm} \normalsize

The importance of the zero-temperature critical magnetic field
$H_{c0}$ for the enhancement of the jumps of the certain
thermodynamic quantities at the equilibrium phase transition point
$T_{\mbox{\scriptsize eq}}$ becomes obvious from
Eqs.~(\ref{eq74}), (\ref{eq75}) and (\ref{eq77}). Eq.~(\ref{eq77})
shows that the order parameter jump $|\psi|_{\mbox{\scriptsize
eq}} = |\psi_0|\varphi_{\mbox{\scriptsize eq}}$  is large for
large values of $H_{c0}$ and $\xi_0$. This is consistent with the
requirement for relatively small values of the GL parameter
$\kappa$, as shown by Eq.~(\ref{eq76}). Therefore, the
unmeasurable ratio $Q/\delta C$ discussed in
Ref.~\cite{Halperin:1974} does not depend on the value of the
critical field $H_{c0}$ but the quantities $Q$ and $\delta C$
themselves, as well as the order parameter jump
$|\psi|_{\mbox{\scriptsize eq}}$ depend essentially on $H_{c0}$.
The values of the reduced order parameter jump
$\varphi_{\mbox{\scriptsize eq}}$ for films of Al, In and W of the
same thickness have the same order of magnitude while the
respective order parameter jump $|\psi|_{\mbox{\scriptsize eq}}$
is one order of magnitude higher for Al than for W, as shown
above. The effect of the critical magnetic field $H_{c0}$ on the
latent heat $Q$ is, however, much stronger and, as is seen from
Table 2, the latent heat $Q$ in W films is very small and can be
neglected while in Al and In films it reaches values, which could
be measured in suitable experiments. This is so because the latent
heat is proportional to the difference $[H_{c0}^2/8\pi \sim
b|\psi_0|^4]$ between the energies of the ground state
(superconducting phase at $T=0$) and the normal state. It is
consistent with the fact that the fluctuation contribution to the
free energy, i.e., the $C-$term in the r.h.s. of Eq.~(\ref{eq71})
is generated by the term of type $|\psi|^2\int d^Dx
\vec{A}^2(\vec{x})$ in the GL free energy. At $T=0$ this free
energy term is also proportional to the mentioned difference
between the free energies of the superconducting ground state and
the normal state.

The shift of the phase transition temperature
$t_{\mbox{\scriptsize eq}} = |(T_{\mbox{\scriptsize
eq}}-T_{c0})|/T_{c0}$, the reduced value
$\varphi_{\mbox{\scriptsize eq}}$ of the equilibrium order
parameter jump $|\psi|_{\mbox{\scriptsize eq}}$, and the latent
heat $Q$ of the equilibrium transition are given for films of
different thicknesses and substances in Table~2. This table shows
that the shift of the phase transition temperature is very small
and can be neglected in all calculations and experiments based on
them. The values for $\varphi_{\mbox{\scriptsize eq}}$ for
different $L_0$ and those for $|\psi_0|$ given in Table~1 confirm
the conclusion, which we have made for films of Al, In, and W with
$L_0 = 0.05\mu$m. The latent heat $Q$ has maximal values for In,
where the critical field is the highest for the considered
materials.

\subsection{Final remarks}

Our analysis shows that the MF studies of the HLM effect have a
well defined domain of validity for both 3D- and quasi-2D
superconductors. Our conclusion is that the MF theory of the
magnetic fluctuations in superconductors and, in particular, the
MF prediction for the fluctuation driven weakly first order phase
transition in a zero external magnetic field in bulk
superconductors~\cite{Halperin:1974} and quasi-2D superconducting
films~\cite{Folk:2001,Shopova2:2003,Shopova3:2003} is reliable and
can be tested by experiments. While the HLM effect in bulk systems
is inobservantly small, in quasi-two dimensional superconductors
this effect is much stronger and may be observed with available
experimental techniques.

Our consideration of quasi-2D superconductors is highly nontrivial
in view of the relevance of the effective Landau parameters
dependence on the thickness of the films, $L_0$. We have justified
this dependence by simple heuristic arguments and by a reliable
consideration of the 2D-3D crossover. In contrast to initial
expectations~\cite{Halperin:1974} that films made of
superconductors with extremely small GL parameter $\kappa$ such as
Al and, in particular, W will be the best candidates for an
experimental search of the HLM effect, our careful analysis
(firstly published in Ref.~\cite{Shopova3:2003}) definitely gives
somewhat different answer. The Al films still remain a good
candidate for transport experiments, through which the jump of the
order parameter at the phase transition point could be measured,
but surprisingly the W films turn out inconvenient for the same
purpose, due to  their very low critical field $H_{c0}$. The
importance of the critical magnetic field $H_{c0}$ for the clearly
manifested first-order phase transition has been established and
discussed, too. Although In has ten times higher GL number
$\kappa$ than Al, the In films can be used on equal footing with
the Al films in experiments intended to prove the order parameter
jump. Here the choice of one of these materials may depend on
other features of experimental convenience. As far as caloric
experiments are concerned, the In films seem to be the best
candidate due to their high latent heat.

As shown in this review experiments, designed to search the HLM
effect, can be performed by both type I and type II
superconductors. In experiments the external magnetic field cannot
be completely eliminated. Then vortex states may occur for $H=
|\vec{H}| > 0$ below $T_c= T_c(H) \leq T_{c0}$ in type II
superconducting films and this will obscure the HLM effect. Note,
that in both type I and type II superconductors
 the external magnetic field $H$ generates additional entropy jump at
the phase transition point $T_c(H)$ and this effect can hardly be
separated from the entropy jump~(\ref{eq74}) caused by the
magnetic fluctuations in the close vicinity of $T_{c0}$.
Therefore, in experiments intended for searching of  HLM effect,
the external magnetic field should be minimized as much as
possible. For quasi-2D superconductors, where the HLM effect is
relatively strong and the latent heat can exceed several ergs, one
should ensure external fields less than 10 Oe, or, in more
reliable experiments, less than 1 Oe~\cite{Shopova5:2003}.

The temperature range around $T_{c0}$, where the HLM effect is
significant, has been estimated in Ref.~\cite{Halperin:1974} to a
few microdegrees, and later, this estimate has been confirmed in a
more accurate way~\cite{Lawrie:1983}. According to our point of
view, the possibilities for the observation of the effect are
related mainly with its magnitude, because the range of
temperatures where it occurs always exists, even in the critical
region of strong $\psi$-fluctuations, as the RG studies indicate
(Sec.~III). The review of the results for Al, In, and W in Sec.~II
shows that the metastability domains (of overheating and
overcooling) are much larger than the Ginzburg critical region.
This result justifies the reliability of the MF treatment. As one
may see from Figs.~2-4 the metastability temperature interval is
relatively larger for smaller values of the GL number $\kappa$,
but in order to ensure large values of some measurable
thermodynamic quantities as, for example, the latent heat $Q$ and
the specific heat capacity $C$ we must choose a material with
large critical field $H_c$. The experimental verification of the
order parameter jump can be made by transport experiments. As we
see from Figs.~2-4 and Table 2, this quantity has maximal values
for W, where the parameter $\kappa$ is very small. Having in mind
also the large metastability regions in this material, one may
conclude that W is a good candidate for a testing  the HLM effect
by transport experiments (measurements of the superconducting
currents), provided some specific disadvantages of this material
(quite low $T_{c0}$, etc.) from the experimental point of view do
not contradict to this our conclusion. Another effect, which is
relevant to the present discussion, is the known variation of the
GL parameter $\kappa$ with the variation of the thickness $L_0$
(see, e.g., Ref.~~\cite{Huebener:2001}. The parameter $\kappa <
1/\sqrt{2}$ of a type-I bulk superconductor may change up to
values corresponding to a type-II superconductor with the decrease
of the thickness $L_0$ below $10^{-7}$m.

In Ref.~\cite{Bettencourt:2003} a very interesting
interrelationship between the first order phase transition and the
vortex fluctuations in type-I 2D superconductors has been
investigated by both analytical and numerical methods. A recent
study~\cite{Abreu1:2004} of HLM effect in superconducting films
has been proven wrong in our comment~\cite{Shopova:2004}; see,
also, the Corrigendum~\cite{Abreu2:2004} to the
work~\cite{Abreu1:2004}. In further papers the same authors and
co-workers confirm~\cite{Abreu3:2004, Abreu4:2004} our results,
published in Ref.~\cite{Folk:2001, Shopova:2002, Shopova2:2003,
Shopova3:2003, Shopova4:2003, Shopova5:2003}, and discussed in
this Section. Another paper has been intended to the treatment of
gauge effects in superconductors with the help of a Gaussian
effective functional~\cite{Marotta:2006}, which gives results
quite near to mean-field ones, and is wrong within the present
discussion.

\section{RG STUDIES AND OTHER TOPICS}

\subsection{Notes about RG}

The RG methods are intended to investigation the strong
fluctuation interactions in the critical region of second order
phase transitions and multi-critical phenomena~\cite{Uzunov:1993,
Zinn-Justin:1993}. In studies of complex systems with several
orderings or with influence of additional gauge fields, as in
Eq.~(1), usually the $\vec{k}-$space RG methods are used because
of their wider applicability. The methods differ from each other
in specific technical details but all of them lead to the same
predictions about the phase transition properties. The common
feature is that the mentioned methods are based on the so-called
loop expansion~\cite{Uzunov:1993, Zinn-Justin:1993, Coleman:1985}.
The RG equations are, in fact, infinite asymptotic series, which
are truncated at one-, two-, or, in rare cases, higher orders in
the loop expansion. The total infinite series can be summed only
in case of trivial (usually exactly solvable) models and in some
very exceptional cases as, for example, the RG series for
interacting real bosons in the quantum limit $T\rightarrow
0$~\cite{Uzunov:1981}.

For the simple $\phi^4-$theory the RG series can be derived and
analyzed to high orders in the loop expansion whereas for more
complex models this can be practically done within the one- and
two-loop approximations. As the RG series are asymptotic,
normally, the one- and two-loop orders give all important features
of the specific system of interest: (i) the presence of stable
fixed points (FP) of RG equations and the associated with them
types of critical behavior, and (ii) the lack of stable FPs and
related conclusions about the lack of standard critical or
multi-critical behavior. The higher orders of the theory are more
relevant to investigations of the asymptotic properties of the
loop series than for  obtaining of qualitatively new critical
behavior, or, other new qualitative characteristics of the system.
The latter are reliably obtained within the one- and two-loop
approximations.

In our consideration we shall give examples of RG equations in the
original variant of the Wilson-Fisher RG approach; see, e.g.,
Ref.~\cite{Uzunov:1993}. It is based on the loop expansion and an
explicit scaling procedure for the wave vectors $(\vec{k} =
e^{-l}\vec{k}^{\prime})$ and the fields $\psi(\vec{k}) = e^{(1 -
\eta_{\psi}/2)l}\psi^{\prime} (\vec{k}^{\prime})$ and
$A_j(\vec{k}) = e^{(1
-\eta_{A}/2)l}A_j^{\prime}(\vec{k}^{\prime})$, where $e^l>1$ is
the so-called re-scaling factor ($l >0$), and $\eta_{\psi}$ and
$\eta_{A}$ are the anomalous dimensions ({\em alias} Fisher
exponents) of the fields $\psi(\vec{k})$ and $A_j(\vec{k})$,
respectively.

\subsection{The order of the phase transition}

In Ref.~\cite{Halperin:1974} the simultaneous effect of $\psi$-
and $\vec{A}$--fluctuations in the fluctuation Hamiltonian
(1) has been investigated by one-loop RG and $\epsilon =
(4-d)$-expansion~\cite{Uzunov:1993} for the general case of
$n/2$-component complex vector field $\psi = \{\psi_{\alpha}; \alpha =
1,...,n/2\}$; note, that this field is equivalent to a
$n$-component real vector field. It has been shown~\cite{Halperin:1974}
 that a stable
FP exists below four dimensions $d = (4-\epsilon)$ only for
symmetry indices $n \geq 365.9$, which are far above the real
symmetry index $n=2$ for usual superconductors and numbers $n=4,6$
corresponding to superconductors with certain unconventional
Cooper pairings~\cite{Uzunov:1990}. For $n < 365.9$ real FP does
not exist at all.
 Besides, the $1/n$-expansion has been
used for the calculation~\cite{Halperin:1974} of the critical
exponents in the so-called large-{\em n} limit (alias ``Hartree limit'')
~\cite{Uzunov:1993,
Zinn-Justin:1993},
where ($n > 365.9$) the phase transition is definitely of second
order.

According to the normal interpretation of RG results, the lack of
stable FP is an indication for a lack of standard second order
phase transition. But usually, additional (non-RG) arguments are
needed to determine the actual order of the phase transition. In
the present case, the lack of stable FP could be a result of the
same mechanism that produces a fluctuation-driven weakly first
order phase transition in MF approximation (Sec.~2). As RG takes
into account both fluctuations of $\psi$ and $\vec{A}$, one can
conclude that the result for the weakly-first order phase
transition in a zero external magnetic field is valid for both
type I and II superconductors. Of course, the size of the effect
(the size of jumps of energy, latent heat, order parameter) will
depend on the specific substance. This path of investigations has
been further developed in the paper~\cite{Chen:1978} and several
problems, opened in the short Ref.~\cite{Halperin:1974}, have been
solved.

In Ref.~\cite{Lovesey:1980}, where both 2D and 3D superconductors
are considered, the HLM effect has been confirmed in one-loop
order (annealed disorder~\cite{Lovesey:1980}) as well as the
availability of second order phase transition for large-{\em n}
 has been proven in another variant of the Hartree
 limit ($n\rightarrow \infty$).

Next, the effect has been confirmed in a RG investigation of the
equation of state below the phase transition point
$T_{c0}$~\cite{Lawrie2:1982, Lawrie3:1982}. These investigations
show that the effect may occur under quite restricted conditions
for the vertex parameters $e$ and $b$ of the functional (1), in
particular, when the effective charge $|e^{\ast}|$ exceeds some
value. A strong restriction on the HLM effect,
 but in the opposite
direction - a requirement for a sufficiently small effective electric charge
$|e^{\ast}|$, has been found also by a Monte Carlo (MC) study~\cite{Dasgupta:1981}
 of a lattice version of the model (1).
Another MC study~\cite{Bartholomew:1983} concludes that at fixed
effective charge ($|e^2| =5$) the HLM effect strongly depends on
the GL parameter $\kappa = \lambda/\xi$: it is well established
for $\kappa \ll 1$, becomes very weak for $\kappa \sim 1$ and
vanishes for large $\kappa$. On the basis of this picture, a
proposal is made~\cite{Bartholomew:1983} that a tricritical point
exists at some ``tricritical'' value of $\kappa$. As $\kappa$ is
related with the charge $e$, this proposal seems to be in
conformity with other investigations. Note, that the first
prediction for change of the phase transition order with the
variation of $\kappa$ and for the possibility of ``tricritical
point`` to occur at some value of $\kappa$, has been made in
Ref.~\cite{Kleinert:1982} on the basis of duality
arguments~\cite{Peshkin:1978} and analytical calculations; see
further development of this approach in
Refs.~\cite{Kiometzis:1994, Kiometzis:1995, Herbut:1997,
Calan:1999}. A further MC study~\cite{Olsson:1998} of type II
superconductors indicates that these systems undergo a second
order phase transition of universality class 3D {\em XY} rather
than a weakly first order phase transition. The two-loop RG
studies~\cite{Folk:1990, Holovatch:1996}
 show that the account of
 higher orders of the loop
expansion allows a lowering of
 the critical value $n_c = 365.9$, but the RG equations still
have no real FP at $D =3$ and $n=2$ (conventional 3D
superconductors), except for a treatment by Pad\`e
analysis~\cite{Holovatch:1996} (see, also, the
review~\cite{Holovatch:1999}). In $2 + \epsilon$ dimensions, the
nonlinear $\sigma$ model exhibits a second-order phase transition
for all values of $n>0$~\cite{Lawrie:1982}. Owing to this result and
more extensive studies~\cite{Freire:2001, Bergerhoff:1996}
 of the dependence of the effective free energy on the
GL parameter $\kappa$, one may suppose that the critical value
$n_c$ vanishes at some dimension $2<d<4$, that is, that a
tricritical point ``located'' at some specific value of $\kappa$
exist where the superconducting phase transition changes its
order; see also, Ref.~\cite{Lawrie:1997}. RG
studies~\cite{Bergerhoff:1996, Herbut:1996} at fixed dimension $d$
have also been carried out with the aim to determine the phase
transition order (see, also. Ref.~\cite{Lawrie:1997} for a
comment).

\subsection{Disorder effects}

\subsubsection{Quenched impurities}

Here we will discuss mainly disorder described by quenched
impurities and inhomogeneities with Gaussian distributions (see,
e.g., Ref.~\cite{Uzunov:1993}). The effects of annealed and
quenched disorder in classical versions of Abelian-Higgs models,
equivalent to the GL model (1) have been investigated by
Hertz~\cite{Hertz:1978, Hertz:1985} in the context of the theory
of spin glasses in case of Dzyaloshinskii-Moriya interaction. The
specific feature of this approach is that the disorder is
associated with the vector gauge field $\vec{A}(\vec{x})$ and can
be used to describe superconductors in random, uncorrelated, or,
which is the same, short-range ($\delta$-) correlated magnetic
fields rather than usual ones. In high energy physics this
approach makes contact with gauge fields, where a Higgs field is
coupled to a random color field. It is natural to expect, as has
been proven~\cite{Lovesey:1980,Hertz:1978}, that the annealed
gauge model will bring a fluctuation-driven first order phase
transition at usual symmetry indices and a second order phase
transition in the Hartree limit ($n \rightarrow \infty$). In
Ref.~\cite{Lovesey:1980} the phase transition in case of quenched
impurities has been predicted to be of second order in a
calculation to one-loop order, in contradiction to the conclusion
for a lack of stable nontrivial FP of the RG equations within the
same one-loop order~\cite{Hertz:1978}. The origin of this
discrepancy is clear from the argument that the lack of stable FP
may result without any change of the Hamiltonian structure,
whereas the conclusion for the second order of phase transition in
Ref.~\cite{Lovesey:1980} has been made, perhaps, incorrectly,
based on the argument of the $|\psi|^4$-term absence for $D=3$ and
the $|\psi|^2\mbox{ln}|\psi|$-term absence for $D=2$.

More usual case of quenched disorder in Abelian-Higgs models has
been considered in Refs.~\cite{Cardy:1982,Uzunov:1983}. This is
the case when the parameter $a$ in (1) has a random part $\varphi(\vec{x})$,
i.e. $a$ is substituted with $\tilde{a} =  a + \varphi(\vec{x})$,
where the Fourier amplitudes $\delta (\vec{k})$ of the {\em random} function
$\varphi(\vec{x})$ obey, for example, the Gaussian distribution
\begin{equation}
\label{eq78} \langle\varphi(\vec{k})\varphi(\vec{k}^{\prime})
\rangle_{rand} = \Delta(k)\delta_{-k,k^{\prime}}.
\end{equation}
Here the parameter $\Delta (k) \geq 0$ is a non-negative function
for any $0\leq k \leq \Lambda$ (hereafter, $\Lambda = 1$). This
disorder is of type random impurities, or, what is the same,
``random critical temperature''~\cite{Uzunov:1993}. In case of the
so-called ``uncorrelated``, or, ($\delta$-) short range correlated
quenched impurities [$\Delta (k) \equiv \Delta = const$], the
system exhibits a spectacular competition between the impurities
and gauge effects. We shall show this by the example of the RG
equations derived in Ref.~\cite{Uzunov:1983}. Using units, in
which $k_B=T_{c0}=\Lambda=1$ and the notations $q_0 = (2e/\hbar
c)$ and $\gamma = \hbar^2/4m$ the RG equations corresponding to
the impurity model given by Eqs. (1) and (78) can be written in
the form
\begin{equation}
\label{eq79}
a^{\prime} = e^{(2-\eta_{\psi})l} \left[ a +
\frac{1}{8\pi^2}\left(\frac{n+2}{2}b
- \Delta \right) \left(\frac{1-e^{-2l}}{2\gamma} - \frac{a}{\gamma^2}l\right)
+ \frac{3}{4\pi}\left(1-e^{-2l}\right)\gamma q^2_0\right],
\end{equation}
\begin{equation}
\label{eq80}
b^{\prime} = e^{(\epsilon-2\eta_{\psi})l}
\left\{b +
\frac{b}{8\pi^2\gamma^2}\left[6\Delta - \frac{1}{2}(n+8)b\right]l -
12\left(\gamma q_0^2\right)^2l\right\},
\end{equation}
\begin{equation}
\label{eq81}
\Delta^{\prime} = e^{(\epsilon-2\eta_{\psi})l}
\left\{\Delta +
\frac{\Delta}{8\pi^2\gamma^2}\left[4\Delta - (n+2)b\right]l\right\},
\end{equation}
\begin{equation}
\label{eq82}
\gamma^{\prime} = e^{-\eta_{\psi}l}\gamma
\left[1 -
\frac{3}{2\pi}q_0^2 l\right],
\end{equation}
\begin{equation}
\label{eq83}
e^{\eta_{A}l}  = 1 +
\frac{n}{12\pi} q_0^2  l,\;\;\;\;
q_0^{\prime} = e^{(\epsilon -\eta_{A})l/2}q_0.
\end{equation}

When the disorder is neglected ($\Delta = 0$; pure superconductors) the
Eqs. (79) - (83) coincide with those
for pure superconductors~\cite{Halperin:1974} and give the results briefly
discussed in Sec.~3.2. For impure superconductors ($\Delta > 0$), these
equations describe a new
stable FP of focal type~\cite{Uzunov:1983} for dimensions $D < 4$,
which exists for symmetry indices $n > 1$ and has a physical
meaning at dimensions $D > D_c(n) =
2(n+36)/(n+18)$~\cite{Uzunov:1983}. In the impure superconductor,
the new focal FP occurs exactly in the domain of symmetry indices
$n < 365.9$, where the HLM analysis~\cite{Halperin:1974} yields a
lack of FP for the respective pure system (for details, see
Ref.~\cite{Uzunov:1983}).
 Having in mind the
asymptotic nature of the $\epsilon$-expansions within the RG
approach~\cite{Uzunov:1993, Lawrie:1987}, one may conclude that
this focal FP governs the critical behavior at the real dimension
$D =3$ in conventional superconductors ($n=2$), although the
direct substitution of $n=2$ in the above expression for the
``lower critical'' dimension $D_c(n)$ yields D$_c(2) = 3.8$. This
problem has been further investigated in Ref.~\cite{Athorne:1985}.
Long-range quenched disorder within the same RG approach to the
model (1) has been considered in Ref.~\cite{Millev:1984}.

\subsubsection{Random field disorder}

The effect of another relevant type of {\em quenched} disorder --
the so-called ``random field disorder''~\cite{Uzunov:1993} -- on
the phase transitions described by the GL functional (1) has been
considered in Ref.~\cite{Busiello:1986}. In order to describe this
type of disorder one should add to the GL model (1) the following
terms~\cite{Busiello:1986}:
\begin{equation}
\label{eq84} -\int d^{D}x\left[h^{\ast}(\vec{x}).\psi(\vec{x}) +
c.c.\right],
\end{equation}
where $h = \left\{h_{\alpha}; \alpha = 1,\dots,n/2\right\}$ is a
random field with a Gaussian distribution of the type
\begin{equation}
\label{eq85}
\langle h^{\ast}_{\alpha}(\vec{k}) h_{\beta}(\vec{k}^{\prime})\rangle_{rand}
= \delta_{\alpha\beta}
\delta_{\vec{k}\vec{k^{\prime}}}h_0^2k^{-\theta}.
\end{equation}

In Eq.~(85), $\theta = 0$ describes the $k\sim 0$ asymptote of
short-range random correlations (in $\vec{x}$-space,
$|\vec{x}-\vec{x^{\prime}}|^{a-d}$; $a \leq 0$), and $0 \leq
\theta \leq d$ describes long-range correlations ($0 \leq \theta =
a < d$); see Refs.~\cite{Uzunov:1993, Uzunov1:1985, Uzunov2:1985}.

  In both short-range and long-range
random field correlations, a new stable FP has been
obtained~\cite{Busiello:1986}. For short-range random correlations
this FP is stable below the upper critical dimension $D_U = 6$ and
for symmetry indices $n> n_c= 10$. This FP describes a quite
specific critical behavior. The situation for long range random
correlations is more
 complicated and here we will advise the reader to follow the discussion in
 Ref.~\cite{Busiello:1986}.

\subsubsection{General conclusion}

The annealed disorder
 leaves the phase transition properties almost identical to those
 in the pure Abelian-Higgs model (1), whereas the quenched disorder
 gives a lower critical values of the symmetry index $n$, below
 which the weakly first order phase transition (HLM effect)
 exists. Thus one may conclude that the quenched disorder acts in a direction
of ``smearing`` the fluctuation-driven first order phase
 transition, i.e. the HLM effect seems to be much weaker in such
 systems.

\subsection{Liquid crystals}

The first theoretical study of the HLM effect on the properties of
the nematic-smectic $A$ phase transition in liquid crystals has
been performed in Refs.~\cite{Lubensky:1974, Lubensky:1978}. The
main result of these investigations is the confirmation of  the
HLM effect in a close vicinity of this liquid crystal phase
transition. Perhaps, because of the liquid crystal anisotropy,
which explicitly enters in the propagator of the gauge
field~\cite{Lubensky:1978}, here the critical value $n_c=38.17$ of
the symmetry index {\em n}, below which the weakly first order
transition occurs, is lower than in pure superconductors $(n_c =
364.9)$  but it is still quite above the real value $n=2$. Thus
the RG predictions in Refs.~\cite{Lubensky:1974, Lubensky:1978} in
seventies of the previous century were in favor of a weakly first
order phase transition whereas the
experimentalists~\cite{Davidov:1979, Litster:1980} at the same
time claimed that the same nematic-smectic {\em A} phase
transition is of second order (for triple and tricritical points
in nematic-smectic-A-smectic-C systems, and the interrelationship
to the present problem, see Ref.~\cite{Huang:1981}), as well as
later papers~\cite{Garland:1994, Durand:1994}. While in 3D
superconductors the HLM effect is very weak and unobservable with
available experimental techniques, the size of the same effect in
3D liquid crystals allows  an observation in case of very precise
experiment and elimination of obscuring effects as, for example,
neighborhood to tricriticality and liquid crystal anisotropy.
Experiments~\cite{Anisimov:1987, Anisimov:1990} consistent with
the HLM effect has been carried out but, as noticed in
Ref~\cite{Bechhoefer:2000}, these experiments are at their
resolution limits and the full implications of the HLM theory
remains to be tested. On the other hand, experiments in
Ref.~\cite{Bechhoefer:2000} demonstrate a discrepancy with the
simple approximation $(\psi=$const) for the smectic-A filed within
the mean-field like approach (Sec.~II).

\subsection{Quantum effects}

 The first account of quantum correlations
(``fluctuations'')~\cite{Uzunov:1993, Hertz:1976, Shopova6:2003}
on the properties of the phase transition to superconducting state
has been performed in Ref.~\cite{Bushev:1980}. It has been
shown~\cite{Bushev:1980} that the dynamical critical exponent $z$,
produced by the quantum correlations at finite temperatures
$(T>0)$, is given by $z = 2+ 18\epsilon/n$. This is purely gauge
result because, as shown in Ref.~\cite{Uzunov:1980}, the critical
dynamics of a superconductor when neglecting the local gauge
effects $(\vec{A}\equiv 0)$ is identical to that described by the
time dependent GL model (TDGL) in case of lack of conservation
laws~\cite{Hohenberg:1977}; a prediction of the latter result has
been given for the first time in Ref.~\cite{Hertz:1976}.

Another paper~\cite{Fisher:1988} on a gauge model of type (1) is
intended to the investigation of the quantum phase transition
($T_{c0} = 0$) in granular superconductors. The simultaneous
effect of the local $U(1)$ symmetry, disorder and quantum
fluctuations has been considered in the
Ref.~\cite{Fisher:1990,Herbut:1998}, where disordered electronic
systems at $T=0$ are studied. The problem was further extended to
the treatment of quantum phase transitions in underdoped
high-temperature superconductors~\cite{Matthey:2006,
Schneider:2006}.

In Ref.~\cite{Continentino:2001} the quantum phase transition at
zero temperature from superconducting state to the state of normal
metal has been discussed within the framework of the HLM
approximation (Sec.~2.2). In the limit $T \rightarrow 0$, a
dimensional crossover of type $d \rightarrow (d + 1)$ occurs due
to the quantum fluctuations of the gauge field $\vec{A}(\tau,
\vec{x})$, where $\tau$ is the imaginary time~\cite{Uzunov:1993,
Shopova6:2003}. Thus the effective free energy of the three
dimensional ``quantum system`` at $T=0$ corresponds to the free
energy of the respective four-dimensional (4D) classical system.
Remember, that the latter is described by Eqs.~(29)-(31) and (49).
Thus the MF like analysis of 3D superconductor in the
zero-temperature limit $T\rightarrow 0$ will give a
fluctuation-driven first order phase transition, described by a
logarithmic term of type $|\psi|^4\mbox{ln}|\psi|$ rather than the
$|\psi|^3$-term known from Ref.~\cite{Halperin:1974}. Accordingly,
in the same zero-temperature limit $(T_c=0)$ the 2D- and  quasi-2D
superconductors will be described by an effective free energy,
containing a $|\psi|^3$-term as is in 3D superconductors with $T_c
\neq 0$. The complete analysis of the quantum phase transitions
($T_c=0$) of type-II superconductors beyond the MF-like
approximation (Sec.~2.2) requires the account of both thermal and
quantum fluctuations of the order parameter field $\psi(\tau,
\vec{x})$ with the help of RG; for example, such fluctuations have
been systematically taken into account in Refs.~\cite{Fisher:1988,
Fisher:1990}.

\subsection{Complex systems}

The RG investigation~\cite{Uzunov1:1981} of systems with
superconducting and other (non-magnetic) orderings shows that the
gauge field $\vec{A}(\vec{x})$ leads to a drastic modification of
the critical behavior in a close vicinity of bicritical and
tetracritical points (for such points, see, e.g.,
Ref.~\cite{Lifshitz:1980, Uzunov:1993}). Near these multicritical
points the superconducting fluctuations can be enhanced by the
fluctuations of another ordering field and, hence, the HLM effect
in such complex systems seems to be stronger than in the more
simple case, discussed so far ~\cite{Uzunov2:1981}. The coupling
between the superconducting field $\psi$ and the magnetization
mode $\vec{M}(\vec{x})$ in models of ferromagnetic superconductors
intended to describe the coexistence of superconductivity and
ferromagnetism in ternary compounds, has been investigated by RG
in Ref.~\cite{Grewe:1979, Grewe:1980}. It has been shown again
that the gauge field $\vec{A}(\vec{x})$ produces a weakly first
order phase transition, enhanced by the $M$-fluctuations. The same
problem has been further opened in Ref.~\cite{Bhattacharyya:1983}
by a lattice version of the GL model and dual
arguments~\cite{Peshkin:1978} with the conclusion that under
certain circumstances the phase transition from the disordered
phase to the phase of coexistence of ferromagnetism and
superconductivity should be, at least in type II superconductors,
of second order in contrast to the RG prediction~\cite{Grewe:1979,
Grewe:1980}. Another investigation of coupled magnetic and
superconducting order parameters, where the gauge effects are also
relevant, is intended to describe a state of vortex solid in
rear-earth and  borocarbide compounds~\cite{Radzihovsky:2001,
Radzihovsky:2005}.

\subsection{Unconventional superconductors}

Another interesting problem is the behavior of unconventional
superconductors~\cite{Uzunov:1990, Blagoeva:1990}, where the field
$\psi$ is a complex vector: $\psi = \{\psi_{\alpha}, \alpha =
2,3\}$, i.e., $n = 4$ or $6$, depending on the type of the
unconventional Cooper pairing. As already mentioned, in RG studies
we may consider an arbitrary value of the symmetry index $n/2 \geq
1$, and to put $n/2 = 2,3$ in our discussion of real
unconventional superconductors. In order to treat these systems we
must add two new terms to the free energy (1), namely,
\begin{equation}
\label{eq86}
\int d^Dx ( \frac{u}{2}|\psi^2|^2 + \frac{v}{2}\sum^{n/2}_{\alpha=1}|
\psi_{\alpha}|^4 ),
\end{equation}
where the parameter $u$ describes the anisotropy of the
unconventional (non-$s$-wave) Cooper pairing and the $v$-term has
been included to represent the crystal anisotropy; as given in
(86) this term represents a cubic crystal anisotropy. It has been
shown~\cite{Millev:1990}, that HLM FP in this case is unstable
towards the parameters $u$ and $v$, describing the anisotropy of
the Cooper pair and the crystal anisotropy, and the new FP points
that appear in the theory are unstable for usual values of the
symmetry index ($n = 4, 6$). Analytical
calculations~\cite{Millev:1990} have been used to demonstrate
 that one of the FP points that
appear in the RG equations, exists for $n  > n_c= 10988$ and is
stable for some values above $n_c$, at least, in the Hartree limit
($n \rightarrow \infty$). Owing to these results, and mainly due
to the very high value of $n_c$,  a conclusion has been made that
the HLM effect should be more pronounced in unconventional
superconductors~\cite{Millev:1990}.

A next step along this path of RG studies has been performed in
Ref.~\cite{Busiello:1991}, where unconventional superconductors
with quenched impurities have been investigated by RG to one-loop
order. For the most interesting cases of $n=4$ and $n=6$, the RG
equations have a focal FP, which is known from the results for
conventional ($u=v \equiv 0$) impure superconductors (Sec.~3.3.1).
However, here this FP is unstable towards the parameters $u$ and
$v$, which means that the superconducting phase transition in a
close (asymptotic) vicinity of the phase transition point is not a
conventional second order phase transition but rather this
transition is a modified version of the fluctuation-driven first
order phase transition, known from Ref.~\cite{Halperin:1974}.

Another interesting problem is related with the gauge effects in ferromagnetic
unconventional superconductors such as UGe$_2$, URhGe, ZrZn$_2$,
where a rich picture of various phase
transitions and (multi)critical points occurs~\cite{Shopova:2005},
including the existence of a completely new type of
finite temperature critical
behavior~\cite{Uzunov1:2006, Uzunov2:2006, Uzunov3:2006} and quantum critical
points. This problem is quite new and since now there are no investigations of
 the effects of the gauge field fluctuations $\delta\vec{A}(\vec{x})$ on the
phase transition properties in ferromagnetic unconventional superconductors.

\subsection{External field effect}

In real experiments the external magnetic field can hardly be
completely eliminated. The study~\cite{Lawrie1:1983} of the HLM
effect in the wider context of nonzero external magnetic field
$H_0$, shows that the weakly first phase transition, discussed so
far can be obtained in Landau gauge from a renormalized theory of
the equation of state in one loop order; see also
Ref.~\cite{Lawrie:1983}. The phase transition at the second
critical magnetic field $H_{c2}$ has been studied by a $\epsilon =
(6-d)$ expansion within the one loop approximation with the
conclusion that the magnetic fluctuation effects should produce a
fluctuation induced first order phase
transition~\cite{Brezin:1985}. Further investigations of this
problem have been performed in the Hartree limit $(n \rightarrow
0)$ ~\cite{Affleck:1985, Radzihovsky:1995}, and in higher orders
of the loop expansion~\cite{Brezin:1990, Mikitik:1992}; for
comments, see Refs.~\cite{Radzihovsky:1996, Herbut1:1996}.

\vspace{0.3cm} {\bf  Acknowledgements}

 Financial support by Grants No. Phys.1507 (NFSR, Sofia). D.I.U.
thanks the hospitality of MPI-PKS (Dresden), where his work on
this article was performed.

\end{document}